\newcount\draft\draft=0			

\pdfoutput=1

\documentclass[10pt,preprint]{sig-alternate}

\usepackage{graphicx,epsfig,sped,times,wide}
\usepackage{algorithm}
\usepackage{clrscode}
\usepackage[noeka]{mathrmletter}
\usepackage{subfigure}
\usepackage{ulem}
\renewcommand{\em}{\normalem}
\usepackage{multirow}
\usepackage[noend]{algorithmic}
\usepackage{epstopdf}
\usepackage{xspace}
\usepackage{mdwlist}
\usepackage{balance}
\usepackage[hyphens]{url}

\newfont{\mycrnotice}{ptmr8t at 7pt}
\newfont{\myconfname}{ptmri8t at 7pt}
\let\crnotice\mycrnotice%
\let\confname\myconfname%

\permission{Permission to make digital or hard copies of all or part of this work for personal or classroom use is granted without fee provided that copies are not made or distributed for profit or commercial advantage and that copies bear this notice and the full citation on the first page. Copyrights for components of this work owned by others than ACM must be honored. Abstracting with credit is permitted. To copy otherwise, or republish, to post on servers or to redistribute to lists, requires prior specific permission and/or a fee. Request permissions from permissions@acm.org.}
\conferenceinfo{SIGCOMM'14,}{August 17--22, 2014, Chicago, IL, USA.}
\copyrightetc{Copyright 2014 ACM \the\acmcopyr}
\crdata{978-1-4503-2836-4/14/08\ ...\$15.00.\\
http://dx.doi.org/10.1145/2619239.2626319}

\newcommand{\name}{PoWiFi\xspace}
\usepackage{color, colortbl}
\usepackage{graphicx}
\DeclareGraphicsExtensions{.pdf,.jpg,.png}
\graphicspath{{.}{./figures/}{./results/}}

\usepackage{listings}
\newcommand{\lil}{\lstinline}

\newenvironment{Itemize}%
{\begin{itemize}%
\setlength{\itemindent}{0em}
\setlength{\itemsep}{0pt}%
\setlength{\topsep}{0pt}%
\setlength{\parsep}{-2pt}
\setlength{\partopsep}{0pt}%
\setlength{\parskip}{0pt}}%
{\end{itemize}}
{\begin{enumerate}%
\setlength{\itemsep}{0pt}%
\setlength{\topsep}{0pt}%
\setlength{\partopsep}{0pt}%
\setlength{\parskip}{0pt}}%
{\end{enumerate}}
\makeatletter
  \newcommand\figcaption{\def\@captype{figure}\caption}
  \newcommand\tabcaption{\def\@captype{table}\caption}
\makeatother

\def\imagetop#1{\vtop{\null\hbox{#1}}}

\newcommand{\xqed}{\nobreak \ifvmode \relax \else
      \ifdim\lastskip<1.5em \hskip-\lastskip
      \hskip1.5em plus0em minus0.5em \fi \nobreak
      \vrule height0.75em width0.5em depth0.25em\fi}

\newcommand{\xref}[1]{\S\ref{#1}}

\newcommand{\figref}[1]{Fig.~\ref{#1}}

\newcommand{\textred}[1]{\textcolor{red}{#1}}
\ifx\noeditingmarks\undefined
   \newcommand{\pgwrapper}[2]{\textred{#1: #2}}
\else
   \newcommand{\pgwrapper}[2]{}
\fi

\usepackage{xxx} 

\makeatletter

\global\def\section{\@startsection {section}{1}{\z@}%
                                   {2ex \@plus 1ex \@minus .1ex}%
                                   {1ex \@plus.2ex}%
                                   {\normalfont\bfseries\scshape\fontsize{11}{13}\selectfont}}
\global\def\subsection{\@startsection{subsection}{2}{\z@}%
                                     {2ex\@plus 1ex \@minus .1ex}%
                                     {1ex \@plus .2ex}%
                                     {\normalfont\bfseries\fontsize{10}{12}\selectfont}}

\def\@maketitle{\newpage
 \null
 \setbox\@acmtitlebox\vbox{%
\baselineskip 20pt
\vskip 2em                   
   \begin{center}
    {\ttlfnt \@title\par}       
    \vskip 1.5em                
{\subttlfnt \the\subtitletext\par}\vskip 1.25em
    {\baselineskip 16pt\aufnt   
     \lineskip .5em             
    \begin{tabular}[t]{c}\hspace{-.15cm}\@author
     \end{tabular}\par}
    \vskip 1.5em               
   \end{center}}
 \dimen0=\ht\@acmtitlebox
 \advance\dimen0 by -10pc\relax 
 \unvbox\@acmtitlebox
 \ifdim\dimen0<0.0pt\relax\vskip-\dimen0\fi}

\begin{document}

\font\ttlfnt=phvb8t at 16pt

\newcommand{\supsym}[1]{\raisebox{6pt}{{\footnotesize #1}}}

\widowpenalty = 10000

\title{Powering the Next Billion Devices with Wi-Fi}
\numberofauthors{1}
\author{Vamsi Talla, Bryce Kellogg, Benjamin Ransford, Saman Naderiparizi,\\
Shyamnath Gollakota and Joshua R. Smith\\
\affaddr{University of Washington}\\
\affaddr{\{vamsit, kellogg, samannp, ransford, gshyam, jrsjrs\}@uw.edu}
}

\maketitle

\begin{sloppypar}

{\bf Abstract --} We present the first {\it power over Wi-Fi} system that delivers power and works with existing Wi-Fi chipsets. Specifically, we show that a ubiquitous piece of wireless communication infrastructure, the Wi-Fi router, can provide far field wireless power without compromising the network's communication performance. Building on our design we prototype, for the first time, battery-free temperature and camera sensors that are powered using Wi-Fi chipsets with ranges of 20 and 17 feet respectively.  We also demonstrate the ability to wirelessly recharge nickel--metal hydride and lithium-ion coin-cell batteries at distances of up to 28 feet. Finally, we deploy our system in six homes in a metropolitan area and show that our design can successfully deliver power via Wi-Fi in real-world network conditions. 

\section{Introduction}
\label{sec:intro}
Starting in the late 19th century, Nikola Tesla dreamed of eliminating wires for both power and communication~\cite{tesla}.  As of the early 21st century, wireless communication is extremely well established---billions of people rely on it every day.  Wireless power however has not been as successful.  In recent years, near-field, short range schemes are gaining traction for certain range-limited applications, like powering implanted medical devices~\cite{freed} and recharging cars~\cite{chargecar} and phones from power delivery mats~\cite{chargemat,magneticmimo,qispec}. More recently researchers have demonstrated the feasibility of powering sensors and devices in the far field using RF signals from TV~\cite{warp,abc} and cellular~\cite{visser2008ambient,parksGSMWARP} base stations. This is exciting, because in addition to enabling power delivery at farther distances, RF signals can be used to simultaneously charge multiple devices due to their broadcast nature.

This paper shows that a ubiquitous piece of wireless communication infrastructure, the Wi-Fi router, can provide far-field wireless power without significantly compromising network performance. This is attractive for three key reasons:
\begin{Itemize}
\item In contrast to TV and cellular transmissions, Wi-Fi is ubiquitous in indoor environments and operates in the unlicensed ISM band where transmissions can be legally modified to deliver power. Repurposing Wi-Fi networks for power delivery can ease the deployment of RF-powered devices without  additional power infrastructure.
\item Wi-Fi uses OFDM, an efficient waveform for power delivery because of its high peak-to-average ratio~\cite{trotter2009power, trotter2010survey}. Given Wi-Fi's economies of scale, Wi-Fi chipsets provide a cheap platform for sending these power-optimized waveforms, enabling efficient power delivery.

\item Sensors and mobile devices are increasingly equipped with $2.4\,\text{GHz}$ antennas for communication via Wi-Fi, Bluetooth or ZigBee. We can, in principle, use the same antenna for both communication and Wi-Fi power harvesting with a negligible footprint on the size of the device. 
\end{Itemize} 
While recent efforts in the RFID community have focused on designing efficient $2.4\,\text{GHz}$ harvesters~\cite{constant2,ambient24ghz}, none of them have demonstrated power delivery using signals from existing Wi-Fi devices. To check if it would ``just work,'' we placed a battery-free temperature sensor equipped with $2.4\,\text{GHz}$ harvesting hardware ten feet from our organization's Wi-Fi router. We found that, over a 24-hour period, the sensor could not reach the minimum voltage of $300\,\text{mV}$ to operate the harvesting hardware.

\begin{figure}[t!]
\vskip -0.15in
\centering
{\footnotesize
\begin{tabular}{c}
	\includegraphics[width=\columnwidth]{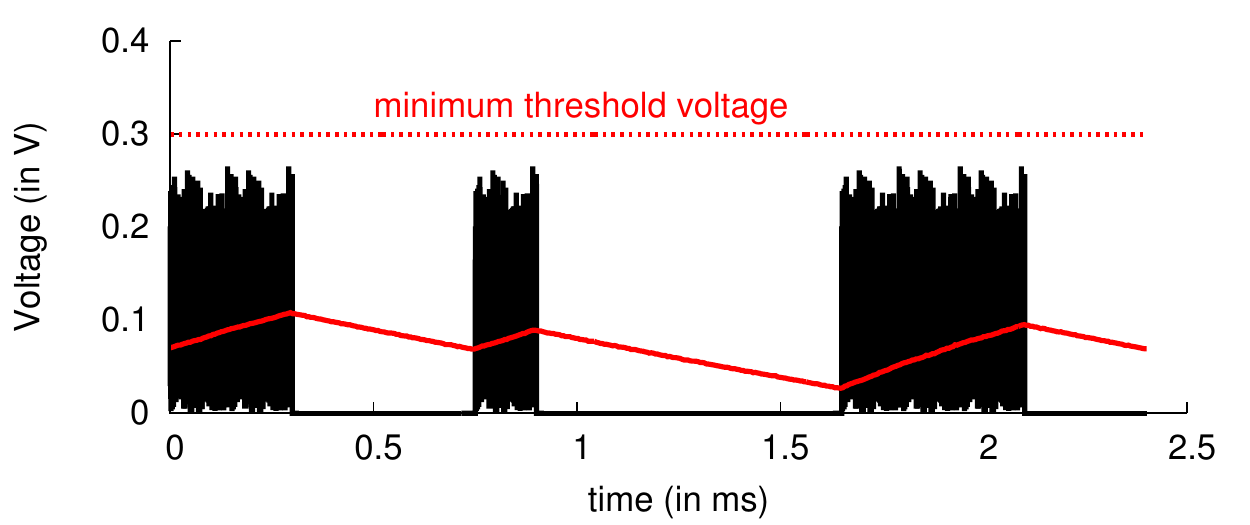}\\

\end{tabular}
}
\vskip -0.15in

\caption{{\bf Key challenge with Wi-Fi power delivery.} While the harvester can gather power during Wi-Fi transmissions, the power leaks during silent periods, limiting Wi-Fi's ability to meet the minimum voltage requirements of the hardware.}

\label{fig:wifi_rectifier}
\vskip -0.15in
\end{figure}

\begin{figure*}[t!]
\vskip -0.1in
\centering
{\footnotesize
\begin{tabular}{cccc}
	\includegraphics[width= 0.23\textwidth]{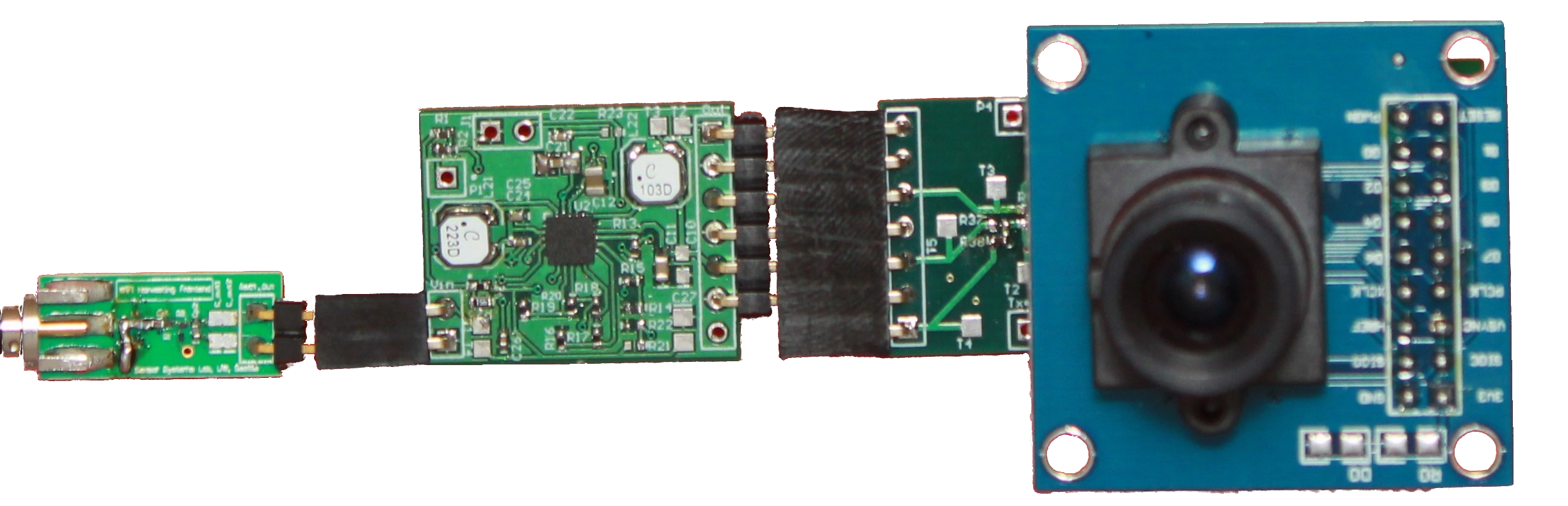} &
	\includegraphics[width= 0.23\textwidth]{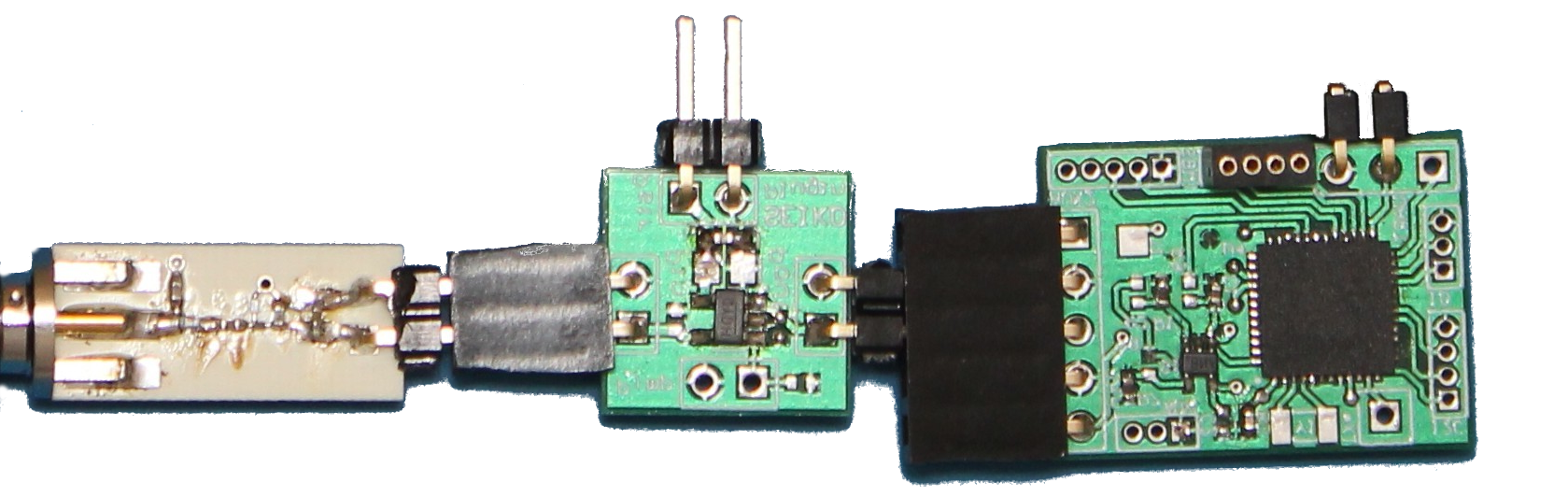} &
	\includegraphics[width= 0.23\textwidth]{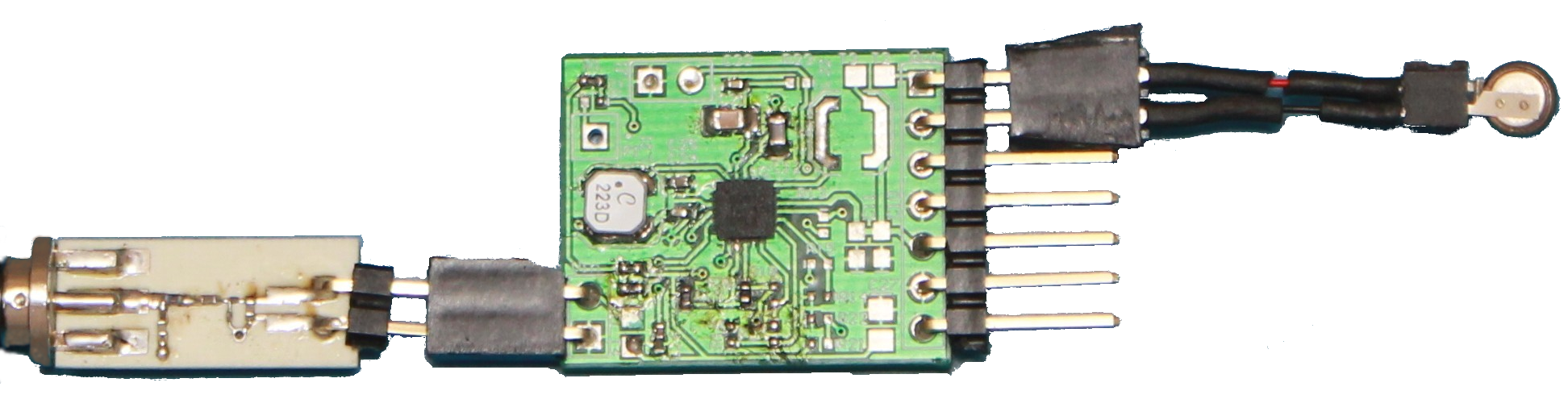} & 
	\includegraphics[width= 0.23\textwidth]{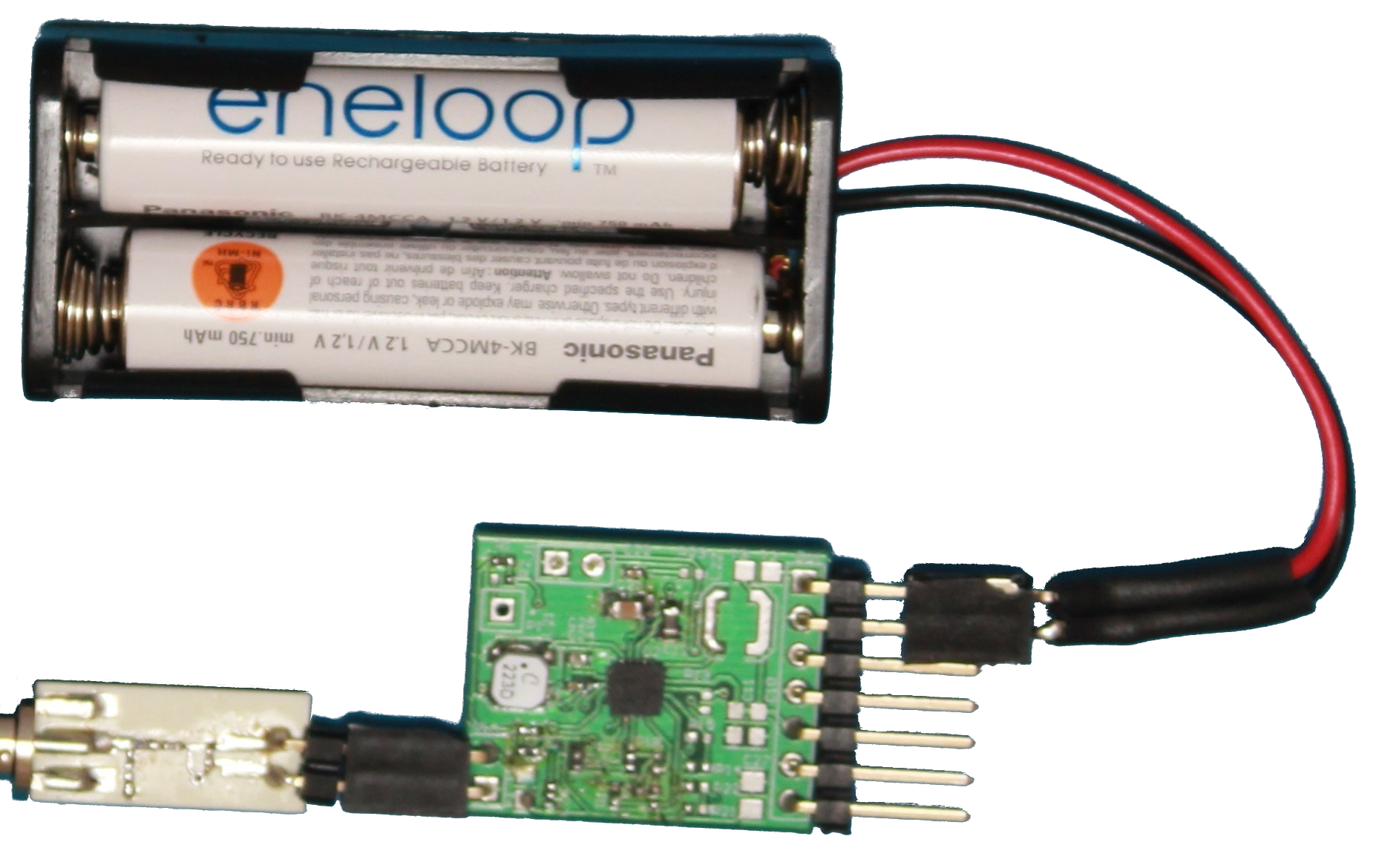} \\

	{(a) Battery Free Camera} & {(b) Temperature Sensor} & {(c) Li-Ion Battery Charger} & {(d) NiMH Battery Charger}
\end{tabular}
}
\vskip -0.1in

\caption{{\bf Prototype hardware demonstrating \name's potential.} The prototypes harvest energy from Wi-Fi signals through a standard $2\,\text{dBi}$ Wi-Fi antenna~\cite{wifi_antenna} (not shown). The low gain antenna ensures that the device is agnostic to the antenna orientation and placement. The prototypes use the harvested energy to (a) capture pictures, (b) measure temperature, and (c)/(d) recharge batteries.}

\label{fig:sensor_imgs}
\vskip -0.09in
\end{figure*}

The key reason for this is the fundamental mismatch between the requirements for power delivery and the Wi-Fi protocol. This is succinctly captured in \figref{fig:wifi_rectifier} which plots the voltage at the harvester in the presence of Wi-Fi transmissions. The figure shows that while the harvester can gather power during Wi-Fi transmissions, the power leaks during silent periods, significantly limiting Wi-Fi's ability to meet the minimum voltage requirement. These silent periods however are inherent to a distributed medium access protocol such as Wi-Fi, where multiple devices share the same wireless medium. A continuous transmission from the router would significantly deteriorate the performance of its own Wi-Fi clients as well as other Wi-Fi networks in the vicinity.

We introduce {\it\name}, the first power over Wi-Fi system that uses existing Wi-Fi chipsets to power energy-harvesting sensors and devices. We achieve this by co-designing harvesting hardware circuits and Wi-Fi router transmissions. At a high level, a router running \name imitates a continuous transmission while minimizing the impact on Wi-Fi performance. To do so, it injects small amounts of superfluous broadcast traffic on multiple Wi-Fi channels (e.g., 1, 6, and 11) such that the cumulative occupancy across the channels is high. We also design a multi-channel harvester that efficiently harvests power across multiple Wi-Fi channels. Because the harvester cannot distinguish between transmissions across channels, it effectively sees a continuous transmission from the router and hence can efficiently harvest power.

The above design has two main components:
\begin{Itemize} 

\item A harvester hardware that efficiently receives power across multiple $2.4\,\text{GHz}$ Wi-Fi channels. Designing this is challenging because a fraction of the incident signal is typically reflected back into the environment and remains unusable because of impedance mismatches in the hardware. While one can minimize these mismatches at a specific frequency tone, achieving well-matched impedances across a range of frequencies is difficult~\cite{fano1950theoretical, matt2, volakis2010planar}. Our approach is to co-design all the harvesting hardware components---rectifier, matching network and DC--DC converter---to achieve reflection losses less than $-10$~dB across the desired $72\,\text{MHz}$ Wi-Fi band.

\item A transmission mechanism at the router that introduces additional \textit{power traffic} on each Wi-Fi channel, while minimizing the impact on Wi-Fi clients and being fair to other Wi-Fi networks. The key insight is that the harvester cannot distinguish between useful client traffic and superfluous power traffic. Thus, the router injects power traffic only when the number of packets queued at the Wi-Fi interface is below a threshold. This minimizes the impact on the associated Wi-Fi clients while effectively providing continuous power delivery to harvesters.  Further, the router transmits power packets at the highest Wi-Fi bit rates. Since higher-rate transmissions occupy the channel for a smaller duration, our scheme achieves per-channel occupancies that are fair to other Wi-Fi networks.

\end{Itemize}

We prototype our router design using Atheros chipsets and build our multi-channel harvester with off-the-shelf analog components. We run extensive experiments to understand the effects of our power traffic on TCP and UDP throughput as well as the page load times of the ten most popular websites in the United States~\cite{alexa-topsites}. Our results show that \name minimizes the effect on Wi-Fi performance while achieving an average cumulative occupancy of 95.4\% across the three $2.4\,\text{GHz}$ Wi-Fi channels.

To demonstrate the potential of our design, we use our harvester to build two battery-free, Wi-Fi--powered sensing systems shown in \figref{fig:sensor_imgs}: a temperature sensor and a camera. The devices use Wi-Fi power to run their sensors and a programmable microcontroller that collects the data and sends it over a UART interface. Our results show that the camera and temperature sensor prototypes can operate battery-free at distances of up to 17 and 20 feet, respectively, from a \name router. As expected, the duty cycle at which these sensor can operate decreases with distance. Further, our sensors can operate in through-the-wall scenarios where they are separated from the router by a wall made of various materials.

Further, we integrate our harvester with $2.4\,\text{V}$ nickel--metal hydride (NiMH) and $3.0\,\text{V}$ lithium-ion (Li-Ion) coin-cell batteries. We then build battery-recharging versions of the above sensors that \name recharges using Wi-Fi. Our results show that the battery-recharging sensors can run energy-neutral operations at distances of up to 28 feet.


Finally, we deploy our router in six homes in a metropolitan area. The occupants of each home used our \name router for their Internet access for 24 hours. Our results show that, even with real-world network conditions, \name efficiently delivers power using Wi-Fi while having a minimal impact on user experience.

\vskip 0.05in\noindent{\bf Contributions.} We make the following contributions:
\begin{Itemize}

\item We introduce \name, a novel system for power delivery using existing Wi-Fi chipsets. We do so without  compromising the Wi-Fi network's communication performance. 
\item To achieve this, we co-design Wi-Fi router transmissions and the harvesting hardware circuits. Our novel multi-channel harvester hardware can efficiently harvest power from multiple $2.4\,\text{GHz}$ Wi-Fi channels.
\item We prototype the first battery-free temperature and camera sensors that are powered using Wi-Fi chipsets. We also demonstrate the feasibility of recharging NiMH and Li-Ion coin-cell batteries using Wi-Fi signals.
\item Finally, we deploy our system in six homes in a metropolitan area and demonstrate its real-world practicality.
\end{Itemize}

\section{Understanding Wi-Fi Power Delivery}
\label{sec:outline}
To understand the ability of a Wi-Fi router to deliver power, we run experiments with our organization's router and a temperature sensor. The router is an Asus RT-AC68U access point operating at 2.437~GHz with a transmit power of 23~dBm on each of its three 4.04~dBi gain antennas. The temperature sensor is battery free and uses our RF harvester to draw power from Wi-Fi signals. A typical RF harvester has to provide a minimum voltage at the sensor or microcontroller to run meaningful operations. This is typically done using a rectifier that converts the carrier signal to DC and a DC--DC converter that increases the voltage level of the DC signal to match the requirements of the sensor or microcontroller. The key limitation in harvesting power is that every DC--DC converter has a minimum input voltage threshold below which it cannot operate. We use the DC--DC converter with the lowest threshold of 300~mV~\cite{seiko}.

We place the sensor ten feet from the router for 24 hours and measure the voltage at the rectifier output throughout our experiments. We also capture the packet transmissions from the router using a high frequency oscilloscope connected through a splitter. Over the tested period, the sensor could not reach the 300~mV threshold. \figref{fig:wifi_rectifier} plots both the packet transmissions and the rectifier voltage. It shows that while the sensor can harvest energy during the Wi-Fi packet transmission, there is no input power during the silent slots. The hardware power leakages during these durations ensure that it does not cross the $300~mV$ threshold.

\figref{fig:wifi_rectifier} is a snapshot of router transmissions during peak network utilization. More generally, the router's channel occupancy was in the 10--40\% range, mostly at the lower end of this range. Note that clients such as smartphones typically transmit at lower power than the router. Our measurements show that, to save energy, smartphones such as Nexus S, Nexus 4 and iPhone 5 reduce their per-packet transmission power to between 0--2~dBm. Thus, efficient power delivery specifically requires high channel occupancies at the router.

\section{\name}\label{sec:design}
\name\ is a novel system that provides power over Wi-Fi using existing Wi-Fi chipsets. At a high level, a \name\ router injects small amounts of unintrusive power traffic on multiple Wi-Fi channels to increase channel occupancy with minimal impact on network performance. We design a multi-channel harvester that cannot distinguish between transmissions on different channels and hence sees an approximation of a high-occupancy router transmission.

\subsection{Multi-Channel Harvester Design}
\label{sec:hardware}

The first goal of our harvester design is to efficiently harvest across multiple 2.4~GHz Wi-Fi channels. A related goal is to achieve good sensitivities across these channels. Sensitivity is the lowest power at which the harvester can boot up and power the sensors and the microcontroller. In theory, one can wait for a long time and harvest enough power to boot up the sensors. In practice, however, because of power leakage, a harvester cannot operate below a minimum power threshold. This is important because the power available at the sensor decreases with the distance from the Wi-Fi router; thus, the harvester's sensitivity determines its operational range.

\begin{figure}[t!]
\centerline{\includegraphics[width=\columnwidth]{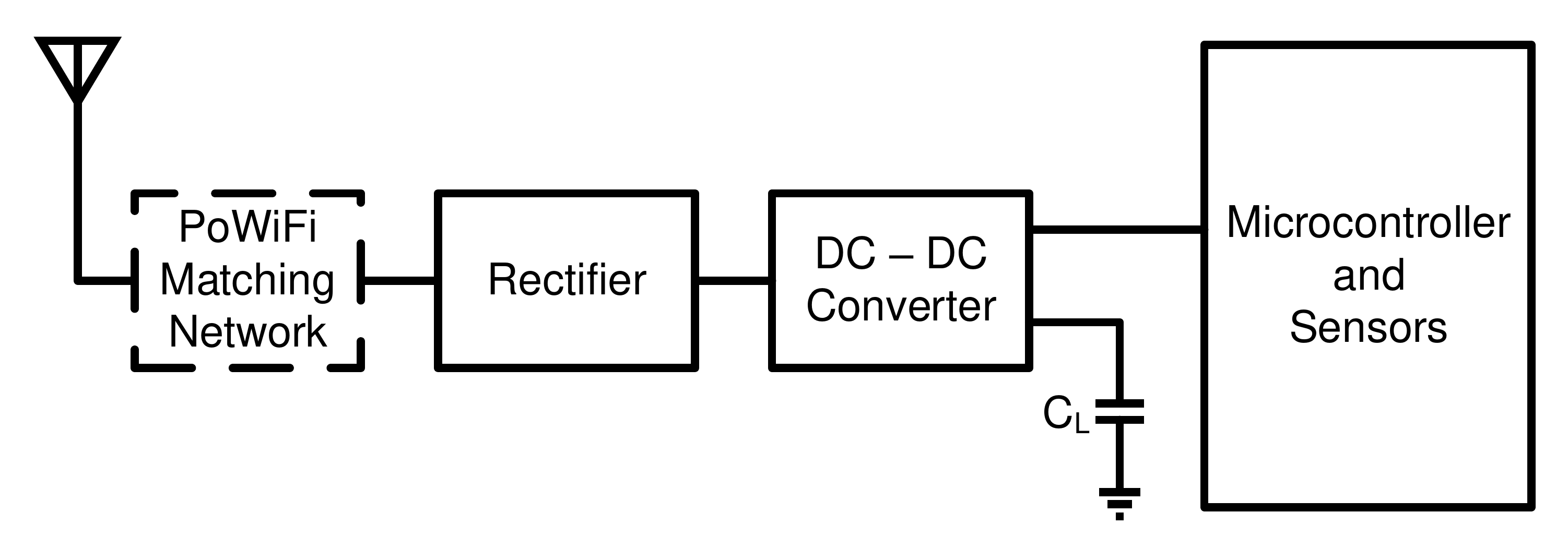}}
\vskip -0.2in
\caption{{\bf RF Harvester Architecture.} An antenna receives RF signals, which a rectifier converts into DC power and feeds into a DC--DC converter that increases the voltage to match the sensor and microcontroller's requirements.}
\label{fig:architecture}
\vskip -0.2in
\end{figure}

\vskip 0.05in\noindent{\bf \textit{Challenge:}} The key challenge is the impedance mismatch between the Wi-Fi antenna and the harvester. To understand this, consider a wave entering a boundary between two different mediums. If the impedance of the two mediums differs, a fraction of the incident energy is reflected. Similarly, when the antenna and the harvester have different impedance values, a fraction of the RF signal is reflected back, reducing the available RF power.

\figref{fig:architecture} shows the architecture of a typical RF harvester. A receiving antenna is followed by a rectifier that converts the 2.4~GHz signal into DC power. This power is fed into a DC--DC converter that increases the voltage of the DC signal to match the voltage requirements of the sensor and microcontroller. The problem is that the rectifier hardware is extremely non-linear with input power, operational frequency and the parameters of the DC--DC converter, making it challenging to achieve good harvester sensitivity and efficiency across the 72~MHz band that spans the three Wi-Fi channels.

\vskip 0.05in\noindent{\bf \textit{Our Approach:}} As shown in \figref{fig:architecture}, we design a matching network to transform the rectifier's impedance to match that of the antenna. This is, however, not straightforward because the rectifier's impedance varies significantly with frequency and is dependent on the DC--DC converter. Our approach is to co-design all the components in the harvester---the matching network, rectifier, and DC--DC converter---to achieve good impedance matching across the 72~MHz Wi-Fi band. Our intuition is that the input of the DC--DC converter affects the input impedance of the rectifier. Thus, if we can co-design the rectifier with the DC--DC converter, we can relax the constraints on the matching network.

\begin{figure}[t!]
\centerline{\includegraphics[width=\columnwidth]{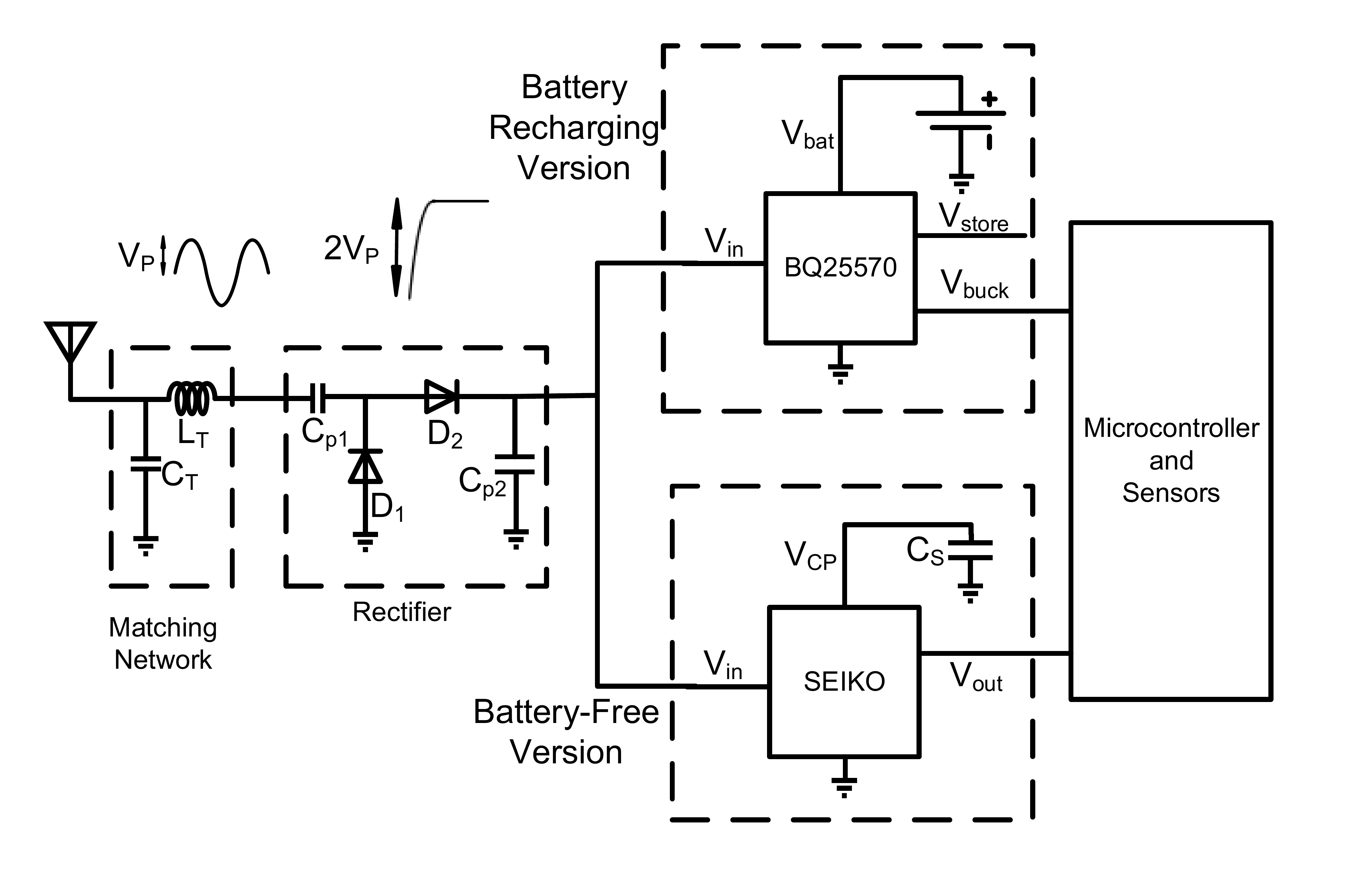}}
\vskip -0.15in
\caption{{\bf \name\ harvester schematic.}  \name\ co-designs the matching network, rectifier, and DC--DC converter to achieve good impedance matching across Wi-Fi bands. The figure shows the optimized DC--DC converters for both battery-free and battery-recharging versions of our harvester.}
\label{fig:schematic}
\vskip -0.25in
\end{figure}

\vskip 0.05in\noindent{\bf \textit{Design Details:}} The rest of the section describes each of the above components---rectifier, DC--DC converter, and matching network---in detail.

\vskip 0.05in\noindent{\it 1) Rectifier Design.} The key design consideration for rectifiers is that DC--DC converters cannot operate below a minimum input voltage. Thus, the rectifier must be designed to maximize its output voltage. \figref{fig:schematic} shows the various components used in our rectifier design. At a high level, our rectifier tracks twice the envelope of the incoming signal and converts it into power. Specifically, it adds the positive and negative cycles of the incoming sinusoidal carrier signal to double the amplitude. To do this, it uses a specific configuration of diodes and capacitors as shown in \figref{fig:schematic}. However, in practice, diodes and capacitors have losses that limit the output voltage of the rectifier. We use SMS7630-061 diodes by Skyworks~\cite{sms7630_061} in ultra-miniature 0201 SMT packages since they have low losses, i.e., loss threshold voltage, low junction capacitance and minimal package parasitics. We also use high--quality-factor, low-loss UHF-rated $10~\text{pF}$ capacitors that minimize losses and maximize the rectifier's efficiency and sensitivity.

\vskip 0.05in\noindent{\it 2) DC--DC converter design.} In our design, a DC--DC converter serves two purposes: i) boost the voltage output of the rectifier to the levels required by the microcontroller and sensors, and ii) make the input impedance of the rectifier less variable across the three Wi-Fi channels. The key challenge is the cold-start problem: in a battery-free design, all the hardware components must boot up from 0~V. Practical DC--DC converters, however, have a nonzero minimum voltage threshold. We use the SZ882 DC--DC converter from Seiko~\cite{seiko}, which is the best in its class: it can start from input voltages as low as 300~mV, which our rectifier can provide, and boost the output on a storage capacitor to 2.4V. Once the 2.4~V threshold is reached, the Seiko charge pump connects the storage capacitor to the output, powering the microcontroller and sensors.

A DC--DC converter can be further optimized while recharging a battery. Specifically, the battery can provide a minimum voltage level and hence the hardware components need not boot up from 0~V. We use the TI bq25570 energy-harvesting chip~\cite{ti_bq25570} that contains a boost converter, a battery charger, voltage monitoring solutions and a buck converter. We connect the rechargeable battery to the battery charging node, $V_{bat}$, of the bq25570. We use the boost as our DC--DC converter to achieve the voltage required to charge the battery. Finally, we leverage the maximum power point tracking (MPPT) mode of the TI chip to tune the input impedance of the DC--DC converter so as to minimize the variation of the rectifier's impedance across Wi-Fi channels. Specifically, we set the buck converter's MPPT reference voltage to 200~mV.

\vskip 0.05in\noindent{\it 3) Matching Network Design:} With our rectifier and DC--DC converter designs, we have relaxed the constraints on the impedance-matching network. The resulting circuit can match impedances between the rectifier and a $50~\Omega$ antenna across Wi-Fi channels, using a single-stage LC matching network. In  LC matching networks, inductors are the primary source of losses. To mitigate this, we use high-frequency inductors in $0402$ footprint which have minimal parasitics and a quality factor of 100 at 2.45~GHz~\cite{coilcraft}. The resulting matching network consumes less board area than traditional transmission lines and distributed-element--based matching networks and can be modified to meet different system parameters without any loss. We use $6.8~\text{nH}$ and $1.5~\text{pF}$ as the LC matching network for our battery-free harvester, and $6.8~\text{nH}$ and $1.3~\text{pF}$ for our battery-recharging harvester.

\subsection{Router Transmission Design}
\label{sec:routerdesign}
Our goal is to maximize power-delivery efficiency that requires maximizing channel occupancy. A na{\"i}ve  solution is to continuously transmit packets at the lowest Wi-Fi bit rate, i.e., 1~Mbps. Since such transmissions occupy the Wi-Fi channel for the longest duration, they maximize the channel occupancy. However, such an approach would significantly deteriorate the performance of Wi-Fi, as our evaluation confirms (see~\xref{sec:traffic_benchmarks}).

Our idea is to instead inject small amounts of traffic on multiple Wi-Fi channels at the router to ensure that cumulative occupancy is high.  The rest of this section first describes how we can inject additional packets while minimizing the effect on Wi-Fi clients and then describes design choices that ensure fairness with other Wi-Fi networks.

The key observation we make is that our harvesting hardware does not decode Wi-Fi signals. As a result, from its perspective, all router transmissions look identical. Thus, it can harvest similar amounts of power from the artificial packets as well as traffic to the Wi-Fi clients and beacon transmissions. We leverage this property to design a system that balances client traffic and additional power traffic.

At a high level, our design injects UDP broadcast packets\footnote{UDP broadcast packets do not require acknowledgments from clients, either at the PHY or the higher layers.} at the highest Wi-Fi bit rate to transmit power on each of the Wi-Fi channels. However, \name drops these broadcast packets when the number of packets in the wireless interface's transmit queue is above a threshold. This ensures that when the router queue has client traffic, we do not add additional packets and hence can minimize the effect on the client delay and throughput.

Specifically, we implement a user-space program that injects 1500-byte UDP broadcast datagrams with a constant inter-packet delay. We use a {\it selective transmission} mechanism that hoists information from the MAC layer to the IP layer. Our mechanism has three main components:
\begin{Itemize}
\item \lil{Power_Socket}:  A standard UDP broadcast socket with the addition of a custom IP option, \lil{IP_Power}, to distinguish its outgoing IP datagrams from other traffic.
\item \lil{Power_MACshim}: A shim interface between the IP stack and the \lil{mac80211} subsystem that enables the IP stack to query the Wi-Fi subsystem for the queue status of individual channels.  On socket creation, the user-space program sets an additional IP option with an integer that uniquely identifies the corresponding wireless interface at the router.
\item \lil{IP_Power}: A mechanism in the IP stack that checks for our power packets on the outgoing IP datagrams and uses our shim interface to decide when to drop the packets.
\end{Itemize}

The decision about dropping packets is performed on a per-packet basis in the packet transmission logic of the IP stack, i.e., \lil{ip_local_out_sk()}, to check whether the pending queue depth  is above a threshold value.  This check is channel specific; it is applied after the kernel has determined a route and therefore an interface for the packet.  If the queue depth is indeed at or above a threshold value, then there are already enough power and Wi-Fi client packets in the queue to maximize channel occupancy.  In this case the router drops the packet before transmitting it and returns the corresponding error code to user space.  On the other hand, if the queue depth is below the threshold value, then \lil{IP_Power} queues the packet for transmission at the MAC layer. We note that in our evaluation, the router is configured to provide Internet connectivity on only one 2.4~GHz Wi-Fi channel. Thus, on the other Wi-Fi channels, there are no client packets in the queue and hence we do not drop any UDP broadcast packets. 

Finally, we summarize some of our key design decisions.

\vskip 0.05in\noindent{\it i) Queue threshold value.} After extensive testing, we set a fixed queue depth threshold of five frames.  Specifically, our tests showed that for thresholds less than five, the occupancy decreases since the queue is repeatedly drained and the user-space program that sends UDP broadcast packets was unable to keep the queue full.  Larger threshold values, on the other hand, required more frequent transmissions, resulting in increased slowdown for client traffic.

\begin{figure}[t]
\centering
\includegraphics[width=\columnwidth]{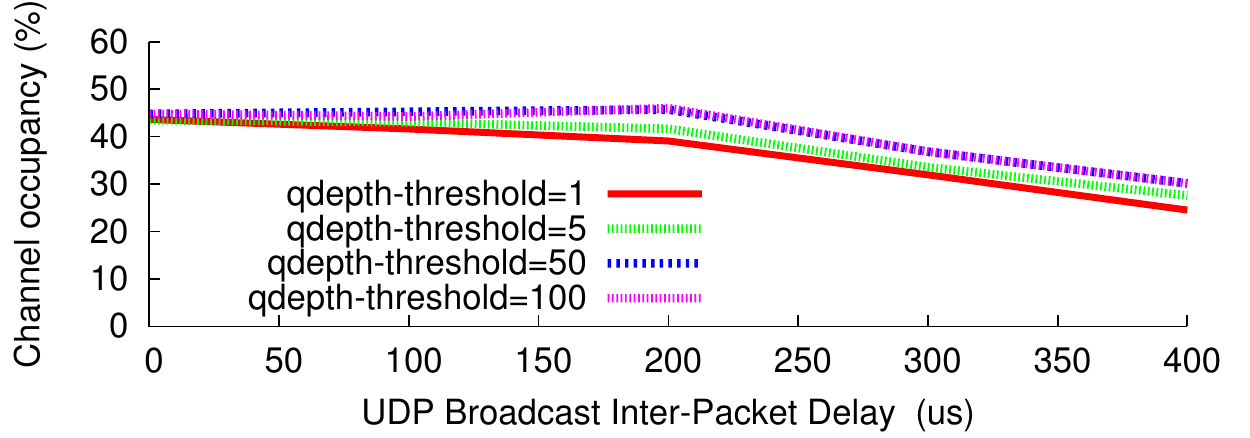}
\vskip -0.15in
\caption{\footnotesize{\bf Effect of inter-packet delay on occupancy.} \textnormal{Results in the absence of client traffic for different queue depth thresholds.}}
\vskip -0.15in
\label{fig:interpkt}
\end{figure}

\vskip 0.05in\noindent{\it ii) UDP broadcast data rate.} If the UDP broadcast rate is high, then frames pile up in the queues and affect the kernel's responsiveness. On the other hand, a low rate significantly reduces the occupancy across Wi-Fi channels. \figref{fig:interpkt} shows the occupancy on a single Wi-Fi channel for different inter-packet delays as well as queue thresholds, in the absence of client traffic. The figure shows that, varying the queue-depth threshold does not significantly affect occupancy in the absence of client traffic as long as inter-packet timing is less than the length of the corresponding frames on the air. Our implementation uses 1500~byte packets transmitted at the highest 802.11g bit rate of 54~Mbps. These packets occupy around 160~us on the wireless channel, and so we pick an inter-packet delay of 100~us to balance occupancy and kernel responsiveness.

\vskip 0.05in\noindent{\it iii) Fairness with other Wi-Fi networks.} \name\ is compliant with the 802.11 MAC protocol to ensure that active Wi-Fi devices get equal access to the wireless channel. In practice, \name\ provides better than equal-share fairness to transmissions from other Wi-Fi devices. Specifically, the UDP broadcast packets are transmitted at the highest Wi-Fi bit rate. These transmissions occupy the channel for a shorter duration than transmissions at lower Wi-Fi bit rates. Thus, for the average transmitter bit rate in the network, we achieve better than equal-share fairness. This is validated in our experiments in~\xref{sec:traffic_benchmarks}.

\section{Evaluation}\label{sec:eval}
We build the rectifiers for our harvester prototypes using 2-layer 20 mils Rodgers 4350 substrate printed circuit boards (PCBs). We use the Rodgers substrate because unlike FR4~\cite{rodgers}, it has low losses at 2.4~GHz and does not degrade the sensitivity and efficiency of our harvester. The DC--DC converter and sensor applications however were built using a 4-layer FR4 substrate PCBs and connected to the harvester using 10 mil headers. The PCBs were designed using Altium design software and were manufactured by Sunstone Circuits. A total of 40 PCBs were ordered at a total cost of \$2500. The off-the-shelf circuit components were hand-soldered on the PCBs and individually tested, requiring a total of 200 person-hours.

We implement a \name\ router using three Atheros AR9580 chipsets that independently run the algorithm in~\xref{sec:routerdesign} on channels 1, 6, and 11 respectively. The chipsets are connected to 6~dBi Wi-Fi antennas via amplifiers; the antennas are separated by 6.5~cm, which is approximately half a wavelength at 2.4~GHz.  Our prototype router provides Internet access to its associated clients on channel 1 via NAT and transmits at 30~dBm, which is within the FCC limit for communication in the ISM band. Since the Atheros chipsets operate independently, the cumulative occupancy across the three Wi-Fi channels can be greater than 100\% in under-utilized networks. One can implement simple algorithms that would scale back the transmission rate for power packets to ensure that the cumulative occupancy remains less that 100\%. We do not currently implement this feature. Note however  that all our sensor and harvester benchmark evaluations were performed in a busy office network where the average cumulative channel occupancy was around 90\%.

\vskip 0.05in\noindent{\it Measuring the router's channel occupancy.} One of our key metrics is the router's channel occupancy that includes both the power packets and packets to its clients. To measure this, we use \lil{aircrack-ng}'s \lil{airmon-ng} tool to add a monitor interface to each of the router's active wireless interfaces. To measure the router's channel occupancy on a specific interface, we start \lil{tcpdump} on the monitor interface to record the radio--tap headers for all frames and their retransmissions. At the end of the duration, we stop \lil{tcpdump} and use \lil{tshark} to extract frames sent by the router, recording the corresponding bitrate and frame size (in bytes). We then compute the average channel occupancy as $\sum_{i \in \mathit{frames}} \frac{ size_i}{rate_i \times total\_duration}$.

\begin{figure*}[t!]
\centering
{\footnotesize
\begin{tabular}{ccc}
	\hspace{-0.18in} \includegraphics[width=2.4in]{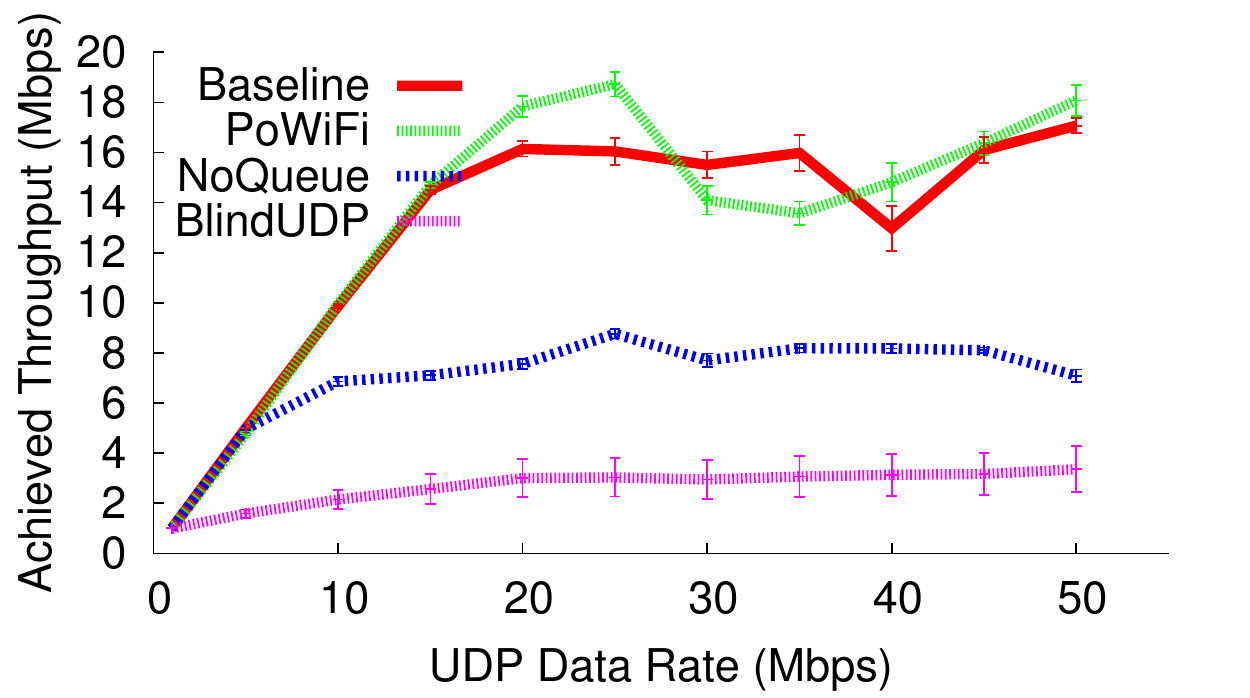} & \hspace{-0.18in} \includegraphics[width=2.4in]{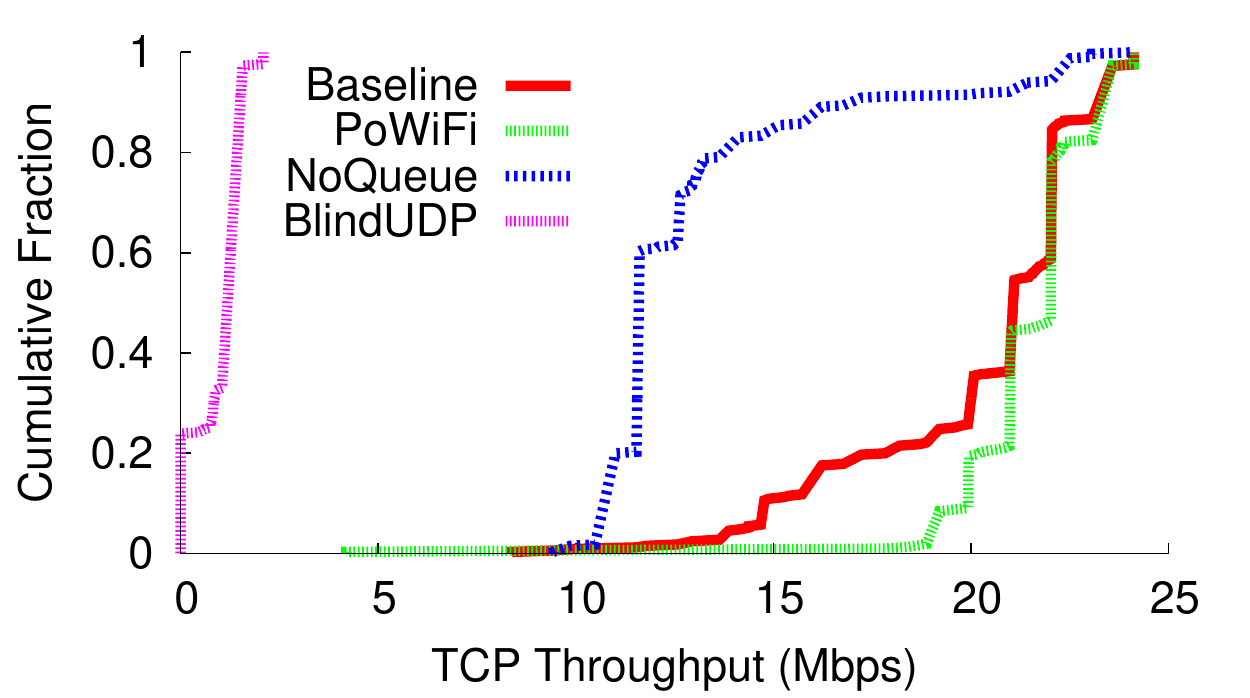} & \hspace{-0.18in} \includegraphics[width=2.4in]{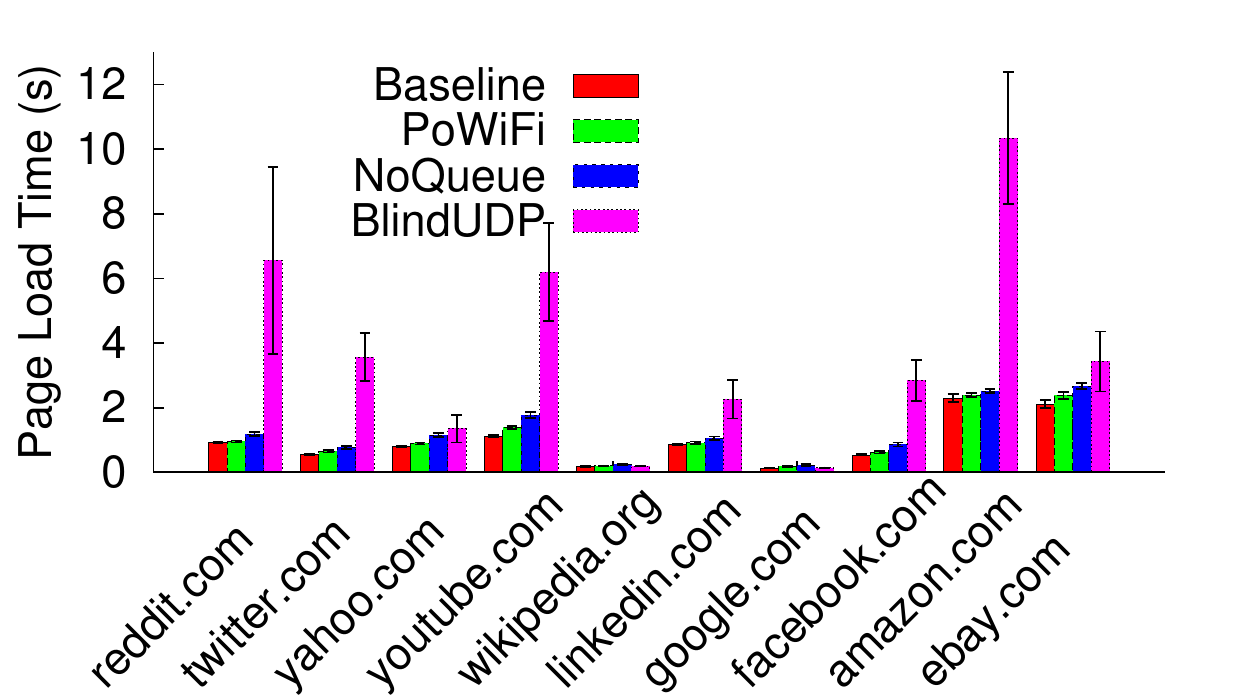}
\end{tabular}
}
\vskip -0.14in
\caption{\footnotesize{\bf Effect on Wi-Fi traffic.} The figures show the effect of various schemes on TCP and UDP throughput as well as the page load times of the top ten websites in the United States~\cite{alexa-topsites}. The plots show that \name\ minimizes its effect on the Wi-Fi traffic.}
\label{fig:traffic}
\end{figure*}

\begin{figure*}[t!]
\vskip -0.1in
\centering
{\footnotesize
\begin{tabular}{ccc}
	\hspace{-0.18in} \includegraphics[width=2.4in]{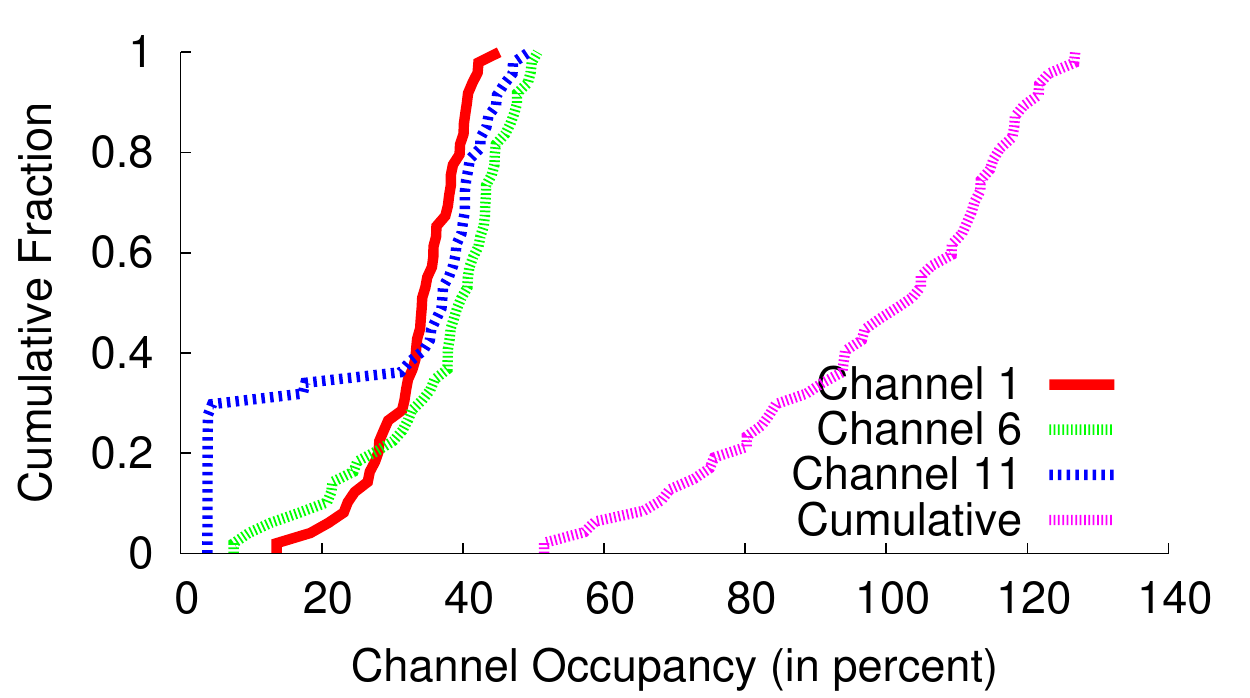} & \hspace{-0.18in} \includegraphics[width=2.4in]{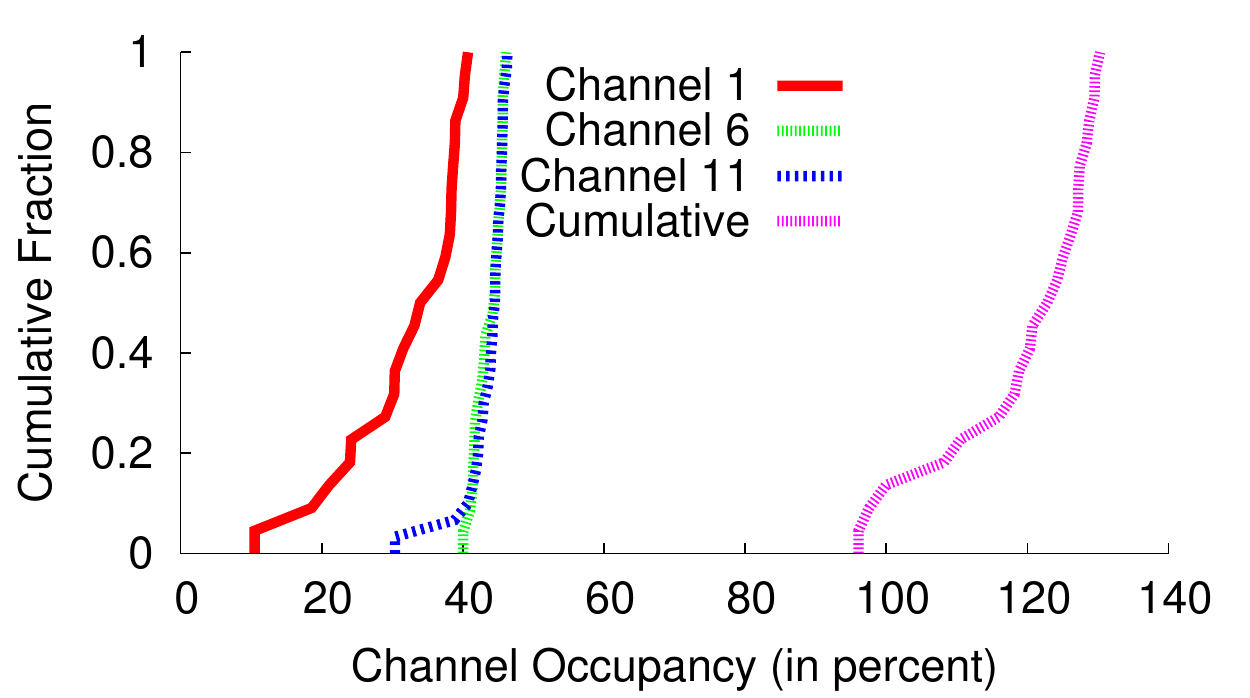} & \hspace{-0.18in}\includegraphics[width=2.4in]{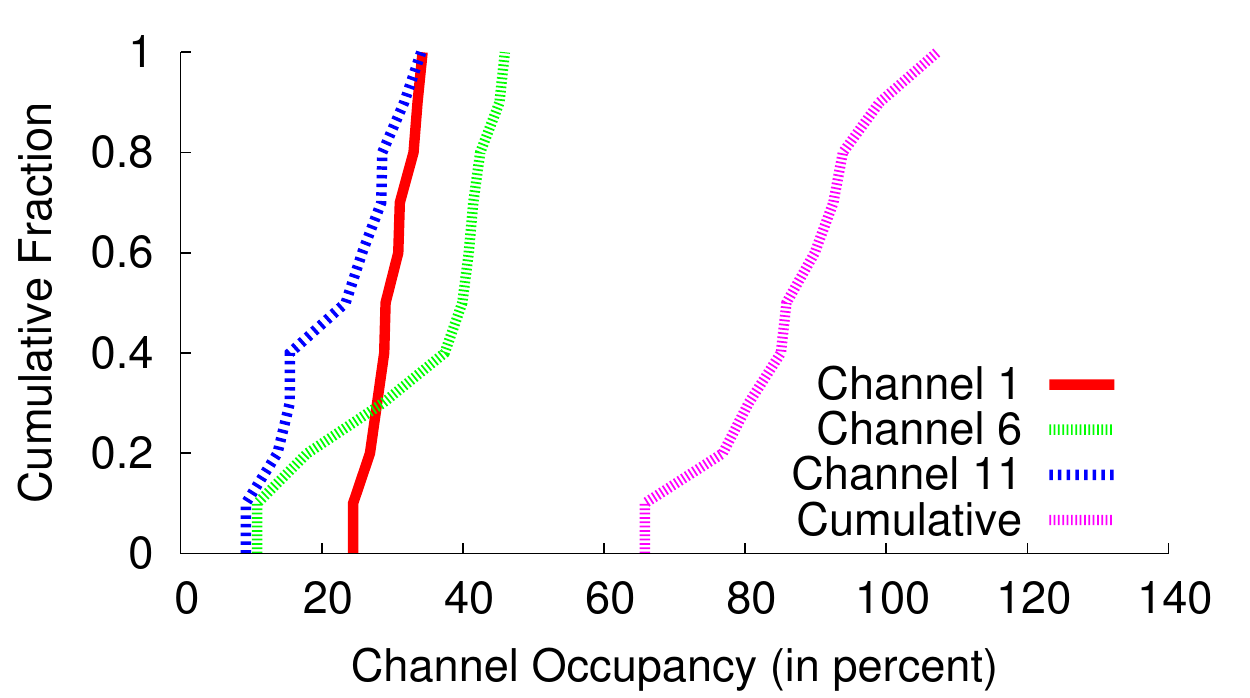}\\
{\it (a) UDP experiments} & {\it (b) TCP experiments} & {\it (c) PLT experiments}
\end{tabular}
}
\vskip -0.14in
\caption{\footnotesize{\bf \name channel occupancies.} The plots show the occupancies with \name\ for the above UDP, TCP, and PLT experiments. }
\label{fig:occupancy}
\end{figure*}

\subsection{Effect on Wi-Fi traffic}\label{sec:traffic_benchmarks}
Our system is designed to provide high cumulative channel occupancies for power delivery while minimizing the effect on Wi-Fi traffic. To evaluate this, we deploy a \name\ router and evaluate its effect on Wi-Fi traffic. We use a Dell Inspiron 1525 laptop with an Atheros chipset as a client associated with our router on channel 1.

We compare four different schemes:

\begin{Itemize}
\item{\it Baseline.} \name is disabled on the router, i.e., the router introduces no extra traffic on any of its interfaces.
\item{\it BlindUDP.} The router transmits UDP broadcast traffic at 1~Mbps so as to maximize its channel occupancy.
\item{\it \name.} The router sends UDP broadcast traffic at 54~Mbps and uses the queue threshold check in~\xref{sec:routerdesign}.
\item{\it NoQueue.} The router sends UDP broadcast traffic at 54~Mbps but disables the queue threshold check.
\end{Itemize}

We evaluate \name with various Wi-Fi traffic patterns and metrics: the throughput of UDP and TCP download traffic, the page load time (PLT) of the ten most popular websites in the United
States~\cite{alexa-topsites}, and traffic on other Wi-Fi networks in the vicinity of our benchmarking network.

\vskip 0.05in\noindent{\it (a) Effect on UDP traffic.} UDP is a common transport protocol used in media applications such as video streaming. We run iperf with UDP traffic to a client seven~feet from the router. The client sets its Wi-Fi bitrate to  $54\,\text{Mbps}$ and runs five sequential copies of iperf, three seconds apart. We repeat the experiments with target UDP data rates between 1 and 50 Mbps, and measure the achieved throughput computed over $500$~ms intervals, with the above schemes. All the experiments are run during a busy weekday in our organization, which has multiple other clients and routers operating on channels 1, 6, and 11.

\figref{fig:traffic}(a) plots the average UDP throughput as a function of the eleven tested UDP data rates.  The figure
shows that {\it BlindUDP} significantly reduces throughput. With {\it NoQueue}, the router's kernel does not prioritize the client's iperf traffic over the power traffic. This results in roughly a halving of the iperf traffic's data rate as the wireless interface is equally shared between the two flows. With \name, however, the client's iperf traffic achieves roughly the same rate as the baseline. This result demonstrates that \name effectively prioritizes client traffic above its power traffic.

For the \name\ experiments above, \figref{fig:occupancy}(a) plots the CDFs of individual channel occupancies on the three Wi-Fi channels. The figure shows that the individual channel occupancies are around 5--50\% across the channels. The mean cumulative occupancy, on the other hand is 97.6\%, demonstrating that \name\ can efficiently deliver power even in the presence of UDP download traffic.

\vskip 0.05in\noindent{\it (b) Effect on TCP traffic.} Next we run experiments with TCP traffic using iperf at the client. The router is configured to run the default Wi-Fi rate adaptation algorithm. We run experiments over a duration of three hours with a total of 30 runs across this duration. In each run, we run five sequential copies of iperf, three seconds apart, and compute the achievable throughput over 500~ms intervals, with all the schemes described above.

\figref{fig:traffic}(b) plots CDFs of the measured throughput values across all the experiments. The plot shows that {\it BlindUDP} significantly degrades TCP throughput. As before, since {\it NoQueue} does not prioritize the client traffic over the power packets, it roughly halves the achievable throughput. \name\ sometimes achieves higher throughput than the baseline. This is because of channel changes that occur during the three-hour experiment duration. The general trend however points to the conclusion that \name\ does not have a noticeable effect on TCP throughput at the client.

\figref{fig:occupancy}(b) plots the CDFs of the channel occupancies for \name\ during the above experiments. The figure shows that \name\ has a mean cumulative occupancy of 100.9\% and hence can efficiently deliver power.

\vskip 0.05in\noindent{\it (c) Effect on PLT.}
We develop a test harness that uses the PhantomJS headless browser~\cite{phantomjs} to download the front pages of the ten most popular websites in the United States~\cite{alexa-topsites} 100 times each. We clear the cache and pause for one second in between page loads.  The  traffic is recorded with \lil{tcpdump} and analyzed offline to determine page load time and channel occupancy. The router uses the default rate adaptation to modify its Wi-Fi bit rate to its clients.  The experiments were performed during a busy weekday in our organization over a two-hour duration.

\figref{fig:traffic}(c) shows that {\it BlindUDP} significantly deteriorates the PLT. This is expected because the 1~Mbps power traffic occupies a much larger fraction of the medium and hence increases packet delays to Wi-Fi clients. {\it NoQueue} improves PLT over {\it BlindUDP}, with an average delay of 294~ms over the baseline. \name\ further minimizes the effect on PLT with a 101~ms delay, averaged across websites. This residual delay is due to the computational overhead of \name\ from the per-packet checks performed by the kernel. This slows down all the processes in the OS and hence results in additional delays. However, increasing processing power and moving these checks to hardware can help further reduce these delays. In our home deployments (\xref{sec:deploy}), the users did not perceive any noticeable effects on their web performance.

For completeness, we plot the CDFs of channel occupancies for \name in \figref{fig:occupancy}(c). The plot shows the same trend as before, with a mean cumulative occupancy of 87.6\%.

\begin{figure}[t]
\centering
\includegraphics[width=\columnwidth]{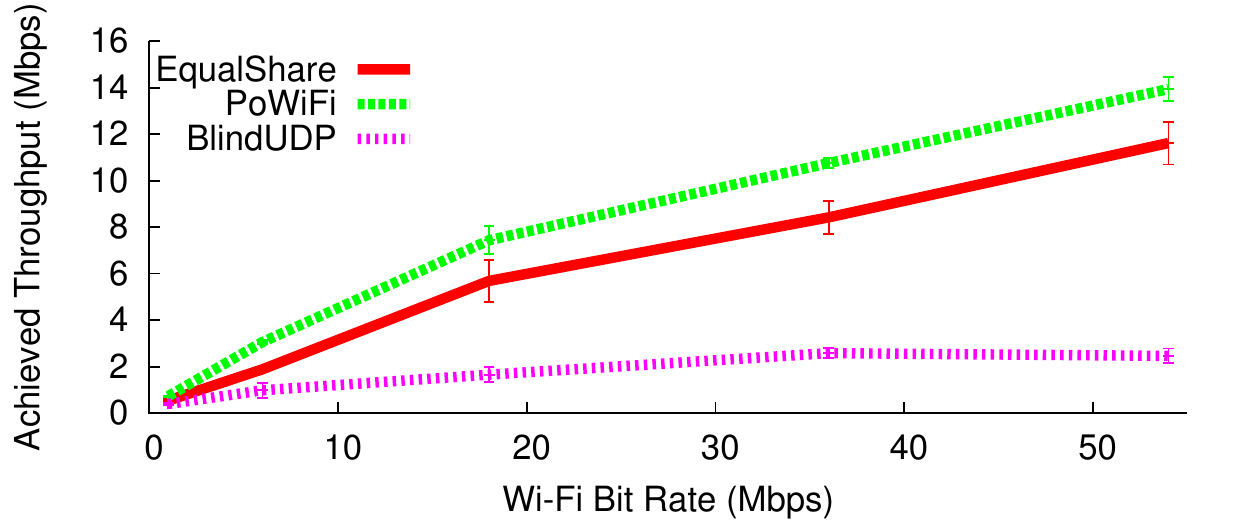}
\vskip -0.1in
\caption{\footnotesize{\bf Effect on neighboring networks.} The figure plots the effect on UDP throughput at various Wi-Fi bit rates on a neighboring network. \name\ provides better than equal-share fairness to other Wi-Fi networks in the vicinity.}
\vskip -0.25in
\label{fig:othernetwork}
\end{figure}

\vskip 0.05in\noindent{\it (d) Effect on neighboring Wi-Fi networks.} \name\ leverages the inherent fairness of the Wi-Fi MAC to ensure that it is fair to other Wi-Fi networks. To evaluate this, we place our \name\ router in the vicinity of a neighboring Wi-Fi router--client pair operating on channel 1. We configure the \name\ router to transmit power packets using our algorithm on all  three channels. We run iperf with UDP traffic on the neighboring router--client pair at the highest data rate and measure the achievable throughput as before. We repeat the experiments for different Wi-Fi bit rates at the neighboring Wi-Fi router--client pair. We compare three schemes: {\it BlindUDP} where our router transmits UDP packets at 1~Mbps, {\it EqualShare} where we set our router to transmit the UDP packets at the same Wi-Fi bit rate as the neighboring router--client pair, and finally \name. {\it EqualShare} provides a baseline when every router in the network gets an equal share of the wireless medium. 

Figure~\ref{fig:othernetwork} shows the throughput for the three schemes, averaged across five runs. As expected, {\it BlindUDP} significantly deteriorates the neighboring router--client performance. Further, this deterioration is more pronounced at the higher Wi-Fi bit rates. With \name, however, the throughput achieved at the neighboring router--client pair is higher than  {\it EqualShare}. This is because \name\ transmits the power packets at 54~Mbps; transmissions at such high Wi-Fi bit rates occupy the channel for a smaller duration than, say, a neighboring router transmitting at 16~Mbps. This property means that \name\ provides better than equal-share fairness to other Wi-Fi networks, preserving their performance. We note that while our experiments are with 802.11g, \name's power packets use the highest bit rate available for Wi-Fi. Thus, the above fairness property would hold true even with 802.11n or other Wi-Fi variants.

\subsection{Evaluating the Harvesting Hardware}
\label{sec:hd_benchmarks}

The harvester's performance is determined by: 1) impedance matching at the antenna interface to maximize the RF energy delivered to the rectifier, and 2) the rectifier's ability to convert RF energy into useful DC power.

\vskip 0.05in\noindent{\it (a) Impedance matching versus frequency.} If the antenna's impedance differs from the harvester's, a portion of the incident RF signal will be reflected back and cannot be converted into DC power. The amount of reflection is determined by the impedance difference, which our matching network aims to minimize across all three Wi-Fi channels. Impedance matching performance is measured using return loss, which is the ratio of reflected power to the incident power.

\begin{figure}[t!]
\vskip -0.1in
\centering
{\footnotesize
\begin{tabular}{cc}
	\hspace{-0.1in}\includegraphics[width=4.3cm]{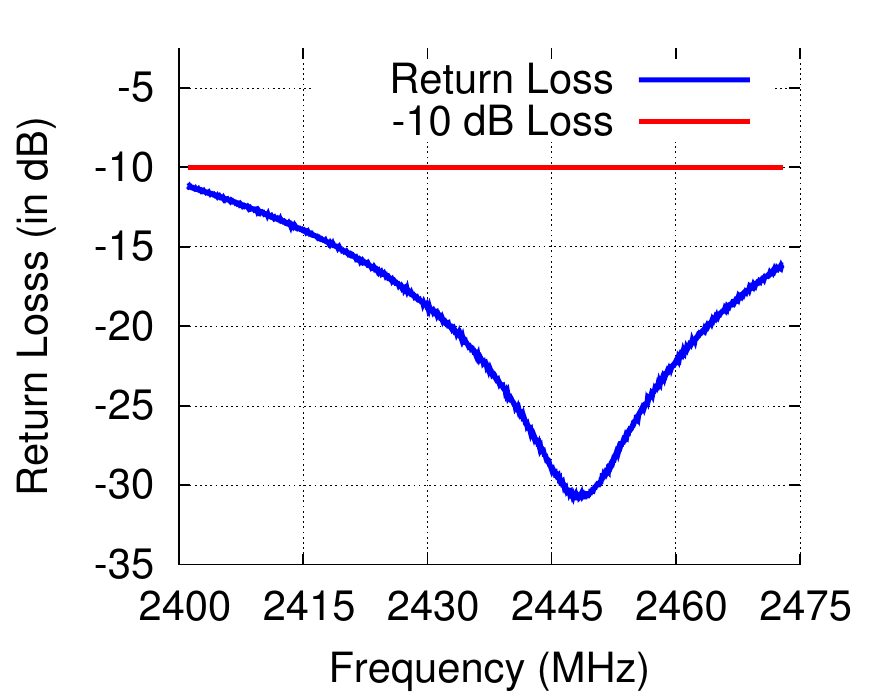} & \hspace{-0.1in}\includegraphics[width=4.3cm]{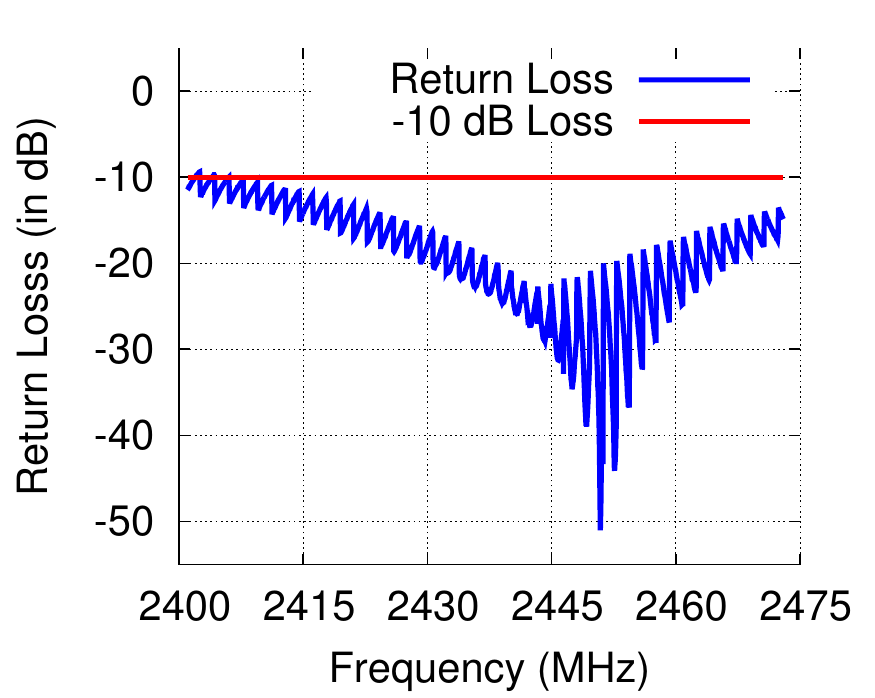} \\
\hspace{-0.1in}{\it (a) Battery-free harvester} &\hspace{-0.1in} {\it (b) Battery-charging harvester}
\end{tabular}
}
\vskip -0.15in
\caption{{\bf Harvester return loss.} This is the ratio of reflected power to the incident power. Across the 2.4~GHz Wi-Fi band, the return loss is less than -10~dB. This translates to less than 0.5~dB of lost power, which is negligible.}
\label{fig:rectifier_rl}
\vskip -0.10in
\end{figure}

We compute the return loss by connecting the harvester to a vector network analyzer that transmits RF signals across the entire Wi-Fi band. We analyze the power reflected at each frequency to compute the return loss.
\figref{fig:rectifier_rl} plots the return loss of the battery-free and battery-charging versions of our harvester. Across 2.401--2.473 GHz, both of our harvesters achieve a return loss of less than $-10\,\text{dB}$, which in most RF circuits and systems is acceptable~\cite{pozar2009microwave}. This translates to less than $0.5\,\text{dB}$ of lost power, which is negligible.

\begin{figure}[t!]
\vskip -0.04in
\centering
{\footnotesize
\begin{tabular}{cc}
	\hspace{-0.1in}\includegraphics[width=4.3cm]{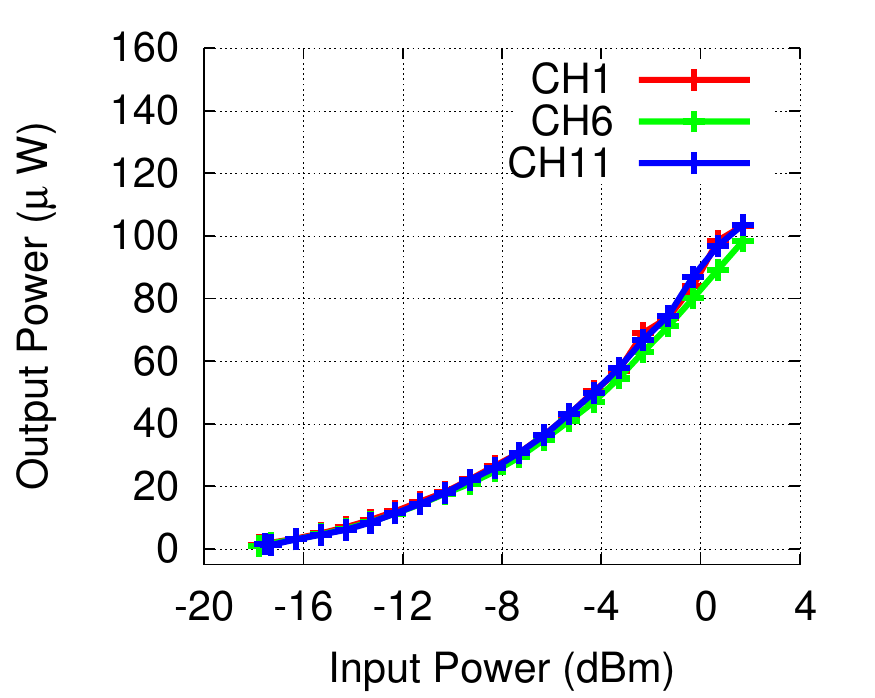} &\hspace{-0.1in} \includegraphics[width=4.3cm]{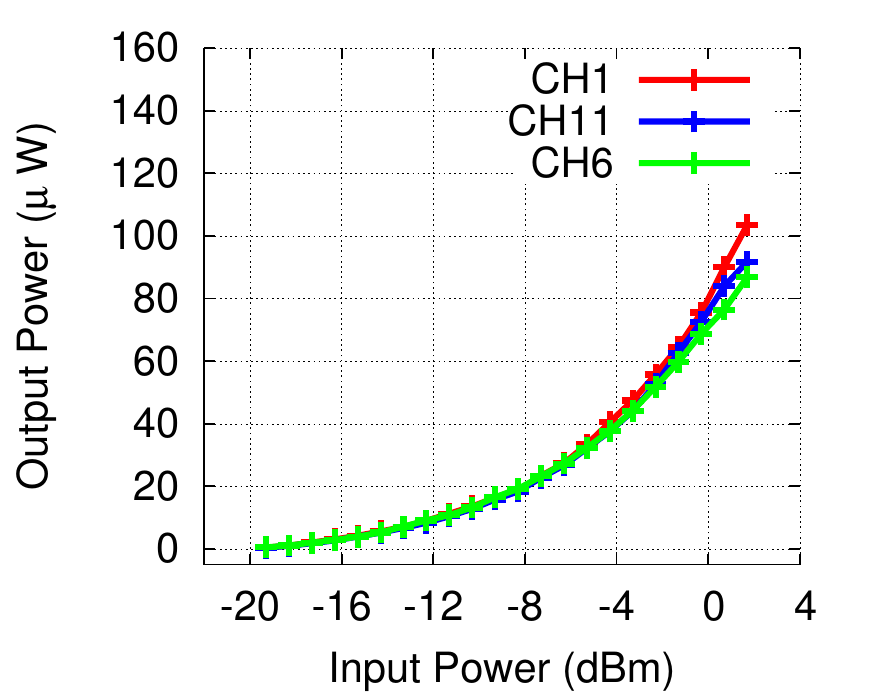} \\
\hspace{-0.1in}{\it (a) Battery-free harvester} &\hspace{-0.1in} {\it (b) Battery-charging harvester}
\end{tabular}
}
\vskip -0.15in
\caption{ {\bf Available output power at the harvester.} The battery charging harvester operates at -19.3 dBm compared to -17.8 dBm for the battery free harvester which results in a higher operating range for the battery charging harvester.}
\label{fig:rectifier_benchmarks}
\vskip -0.09in
\end{figure}

\vskip 0.05in\noindent{\it (b) Available power at the rectifier output.} The rectifier converts the RF signals at the harvester into DC output voltage. This conversion is typically low due to the inherent nonlinearities and threshold voltage drop of diodes. To measure the available power, we use a cable to connect our hardware to the output of a Wi-Fi transmitter. We vary the output power and the operational frequency of the transmitter and measure the power available at the rectifier's output.

\figref{fig:rectifier_benchmarks} shows the output power at the rectifier as a function of the input RF power. The results are plotted for both our battery-free as well as battery-charging harvesters, across the three Wi-Fi channels. The plots show the following:
\begin{Itemize}
\item The harvester's available output power scales with the input power. This means that as a harvesting sensor moves closer to the router, it can operate at a higher duty cycle.
\item The battery-charging harvester operates down to -19.3~dBm, compared to -17.8~dBm for the battery-free harvester. This is because the battery-charging harvester does not have the cold start limitation. Specifically, a battery-free harvester has to start all its hardware components from cold start (0~V). In contrast, a battery-charging harvester can use the connected battery to provide a non-zero voltage value, allowing for greater sensitivities. 
\item Our harvesters perform efficiently across Wi-Fi channels 1, 6 and 11. This is a result of our optimized multi-channel harvester design that ensures efficient power harvesting.
\end{Itemize}

\section{Sensor Applications}
\label{sec:sensor}

We integrate our harvesters with sensors at two ends of the energy-consumption spectrum: a temperature sensor and a camera. We build both battery-free and battery-recharging versions of each sensor.

\subsection{Wi-Fi powered Temperature Sensor}
\label{sec:sensor_temp}
The battery-free temperature sensor uses our harvester to power an LMT84 temperature sensor~\cite{lmt84} and an MSP430FR5969 microcontroller to read and transmit sensor data~\cite{MSP430FR5969_datasheet}. The MSP430FR5969 requires at least 1.9~V to run at 1~MHz and boots in less than 2~ms. When the storage capacitor's voltage reaches 2.4~V, the microcontroller boots, samples the temperature sensor, and transmits the reading through a UART port. The microcontroller's firmware is optimized for power: the entire measurement and data-transmission operation uses only $2.77~\mu\text{J}$.

The battery-recharging sensor, on the other hand, consists of our rectifier followed by the TI bq25570 power-management chip~\cite{ti_bq25570} to wirelessly recharge two AAA 750~mAh low discharge current NiMH battery at 2.4~V~\cite{nimh_battery}. We connect the batteries to the TI chip's $V_{bat}$ node. The temperature sensor and microcontroller are powered from the $V_{store}$ node of the chip, which is internally connected to the NiMH battery. The energy per operation is $2.77~\mu\text{J}$ as above.

\begin{figure}[t]
\centering
{\footnotesize
	\includegraphics[width=\columnwidth]{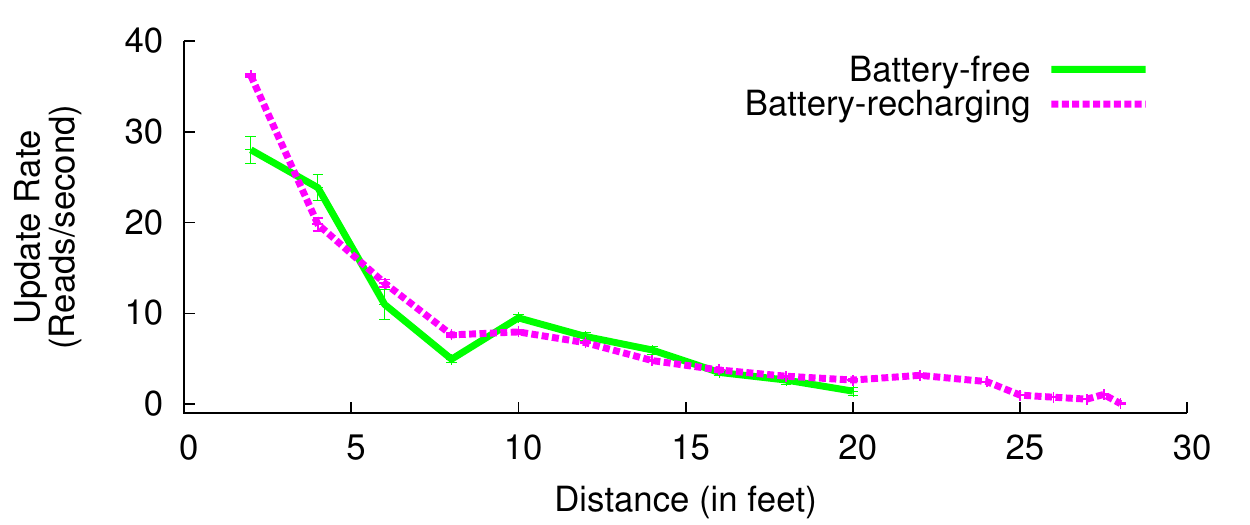}
}
\vskip -0.25in
\caption{{\bf Update rate of temperature sensors.} 
The battery-free sensor can operate up to 20 feet and the battery-recharging sensor can operate in an energy-neutral manner up to 28 feet.}

\label{fig:tempsensor_benchmarks}
\end{figure}

\vskip 0.05in\noindent{\it Experiments.} We evaluate the effect of distance on the update rate of the temperature sensor. Specifically, we use a \name router and place both the battery-recharging and battery-free sensor at increasing distances. In the battery-free case, we measure the update rate by computing the time between successive sensor readings. In the battery-operated case, we measure the battery voltage and the charge current flowing into it from the harvester. Since, each temperature sensor measurement and data transmission takes $2.77\,\mu\text{J}$, we compute the ratio of the incoming power to this value to ascertain the update rate of the sensor for energy-neutral operation. The average occupancy across the Wi-Fi channels in our experiments was 91.3\%.

\vskip 0.05in\noindent{\it Results.} Fig.~\ref{fig:tempsensor_benchmarks} plots the results for both our sensors. The update rates decrease with the distance from the router. This is a result of less power being harvested and agrees with the harvester benchmarks in~\xref{sec:hd_benchmarks}. At closer distances, both harvesters have similar update rates. Beyond 15~feet, however, the battery-powered sensor, optimized for lower input power, has a better update rate and extended operational range: it can operate up to 20 feet from the router. The battery-recharging sensor can operate in an energy-neutral manner to greater distances of up to 28 feet.

\begin{figure}[t]
\vskip -0.1in
\centering
{\footnotesize
	\includegraphics[width=\columnwidth]{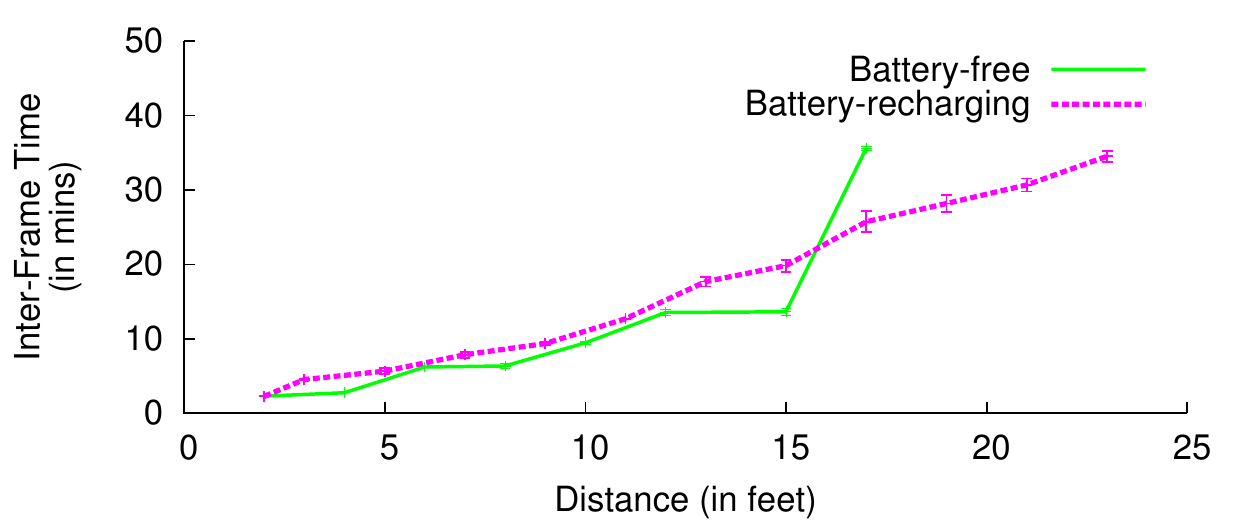}
}
\vskip -0.25in
\caption{{\bf Camera prototype results.} The battery-free camera operates up to 17 feet and the battery-recharging camera has a range of 23 feet for energy-neutral operations. This enables applications where low-rate cameras can be left in hard-to-reach places such as walls, attics, and sewers for leakage and structural integrity detection, without the need for replacing batteries. }
\label{fig:camera_benchmarks}
\end{figure}

\begin{figure}[t]
	\vskip -0.1in
	\centering
	{\footnotesize
		\begin{tabular}{cc}
			\hspace{-0.1in} \imagetop{\includegraphics[width= 2.3cm]{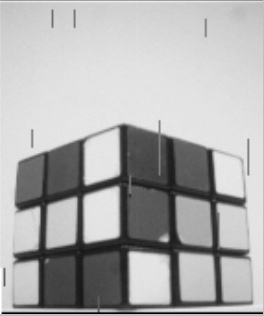}} & \imagetop{\includegraphics[width=6.0cm]{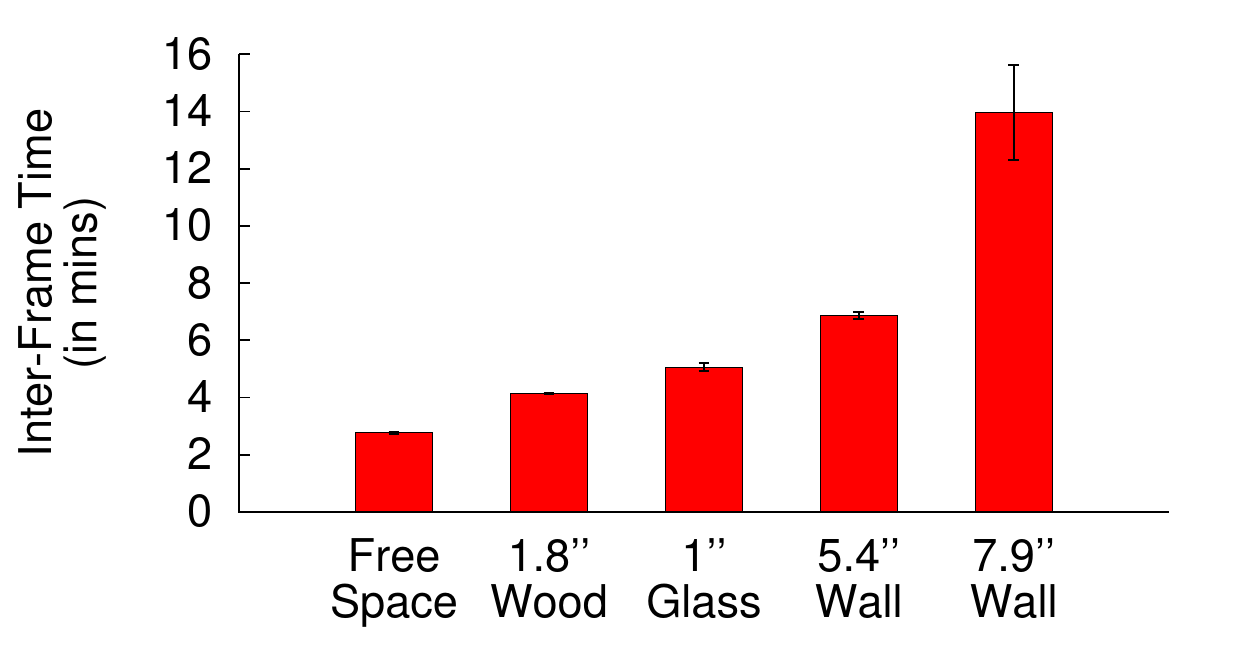}}\\
		\end{tabular}
	}
	\vskip -0.15in

	\caption{{\bf Battery-free camera in through-the-wall scenarios.}
		The figure on the left is a picture of a Rubik's cube taken with our camera prototype. 
		The plot shows the inter-frame time with different wall materials at a five
		feet distance from the router.}

	\label{fig:throughwall}
\end{figure}

\subsection{Wi-Fi powered Camera}
\label{sec:sensor_camera}
We use OV7670, a low-power VGA image sensor from Omnivision~\cite{ov7670}, and interface it with an MSP430FR5969 microcontroller. The image sensor requires a minimum voltage of 2.4~V and consumes 60~$mW$ in active mode operation. We program the sensor to operate in gray-scale QCIF image capture mode with $176\times 144$ frame resolution. The microcontroller initializes and provides timing signals to the image sensor. We transfer the sensor data at 48Mbps and store it on the 64 KB non-volatile FRAM on the microcontroller. We optimize our firmware code for power and achieve a per-image capture energy of 10.4~$mJ$.

On our battery-free camera, we use an ultra-low leakage AVX BestCap $6.8~mF$ super-capacitor as the storage element~\cite{avx_supercap}. The image sensor and microcontroller are powered by the buck converter of the TI bq25570 chip, which provides 2.55~V regulated output voltage. The TI chip activates the buck converter when the super-capacitor voltage reaches 3.1~V and is active until it discharges to 2.4~V. Our battery-recharging camera consists of the same hardware as before, but uses our wirelessly rechargeable 1~mAh lithium-ion coin-cell battery at 3.0~V~\cite{li_battery}.

\vskip 0.05in\noindent{\it Experiments 1.} We evaluate the time between frames as a function of distance for both our camera prototypes. As before, we use a \name\ router---the observed average cumulative occupancy of 90.9\% across experiments. At each distance from the router, we wait for the camera to take at least six frames and measure the time interval between consecutive frames. For the battery-recharging camera, as before, we ascertain the inter-frame duration for an energy-neutral image capture.

\vskip 0.05in\noindent{\it Results 1.} Fig.~\ref{fig:camera_benchmarks} shows that the battery-free camera can operate up to 17 feet from the router, with an image capture every 35 minutes. On the other hand, the battery-recharging camera has an extended range of 23 feet with an image capture every 34.5 minutes in an energy-neutral manner. Both the sensors have a similar image capture rate up to 15 feet from the router. We also note that Fig.~\ref{fig:camera_benchmarks} limits the range to 23~feet to focus on the smaller values. Our experiments, however, show that the battery-recharging camera can operate up to 26.5~feet with an image capture every 2.6~hours.

A key question the reader should ask is: {\it would cameras with such low image-capture rate be useful in practice?} Taking a picture periodically, as above, is an artificial construct of our experiment. In practice, we could integrate our camera with motion-detection sensors that consume orders of magnitude lower power~\cite{allsee} and turn on the camera only when motion is detected. Another application is to use these cameras in hard-to-reach places such as walls, attics, pipes and sewers for leakage and structural-integrity detection. In these scenarios, replacing batteries can be cumbersome, and our low rate camera sensor would be an effective solution.

\vskip 0.05in\noindent{\it Experiments 2.} Motivated by the above applications, we next evaluate our camera in through-the-wall scenarios.  We place our \name\ router next to a wall and place our battery-free camera prototype 5~feet away on the other side of the wall. We experiment with walls of four different materials: a double-pane glass wall of thickness one inch, a wooden door with thickness 1.8 inches, a hollow wall with thickness 5.4 inches, and finally a double sheet-rock (plus insulation) wall with a thickness of 7.9 inches.

\vskip 0.05in\noindent{\it Results 2.} Fig.~\ref{fig:throughwall} shows the mean time between frames, averaged over five frames, as a function of the material. The plot shows that as the material absorbs more signals (e.g., double sheet-rock versus glass), the time between frames increases. However, the key conclusion is that \name\ can power battery-free cameras through walls and hence can enable applications where the cameras can be left in hard-to-reach places such as walls, attics, and sewers, without the need for replacing batteries.

\section{Home Deployment Study}
\label{sec:deploy}

\definecolor{Gray}{gray}{0.85}

{\footnotesize
	\begin{table}[t!]
		\vskip -0.1in
		{
			\centering
			\begin{center}
				\caption{\small{\bf Summary of our home deployment}}
				\label{tab:home}
				\begin{tabular}{|c|cccccc|}
					\hline
					\rowcolor{black}
					{\bf\color{white}Home \# } & {\bf\color{white}1} & {\bf\color{white}2} &
					{\bf\color{white}3} & {\bf\color{white}4} & {\bf\color{white}5} & {\bf\color{white}6} \\
					\hline
					\rowcolor{Gray}
					{\bf Users} & 2  & 1 & 3& 2 & 1 & 3 \\
					\hline
					\rowcolor{Gray}
					{\bf Devices} & 6 & 1& 6 & 4 & 2 & 6\\
					\hline
					\rowcolor{Gray}
					{\bf Neighboring APs} & 17&4& 10&15&24&16\\
					\hline
				\end{tabular}
			\end{center}
		}
		\vskip -0.15in
	\end{table}
}

\begin{figure*}[t!]
\vskip -0.1in
\centering
{\footnotesize
\begin{tabular}{ccc}
	\hspace{-0.25in} \includegraphics[width=2.5in]{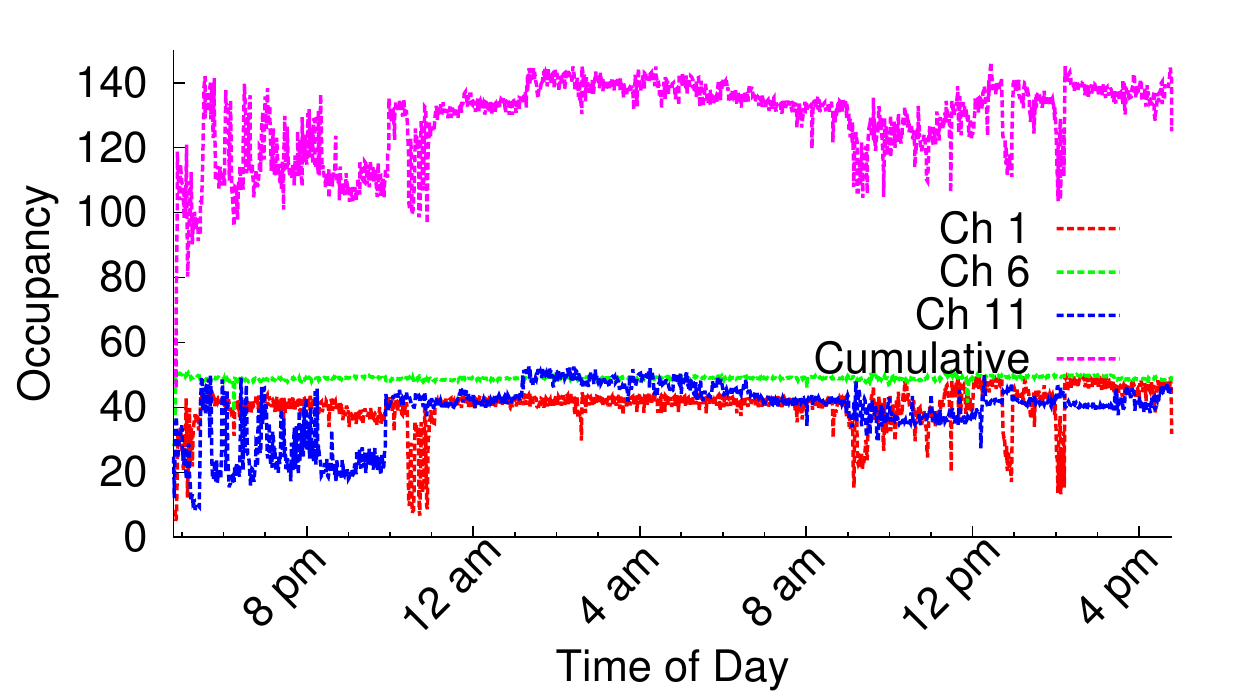} & \hspace{-0.25in} \includegraphics[width=2.5in]{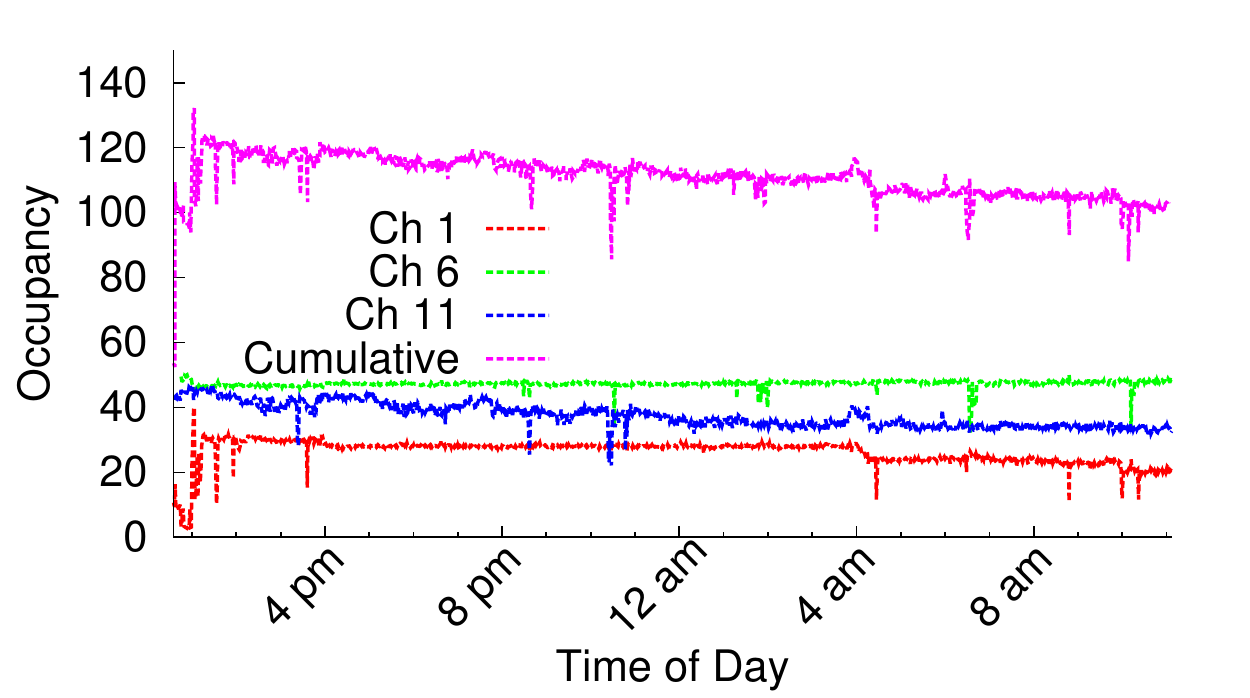} & \hspace{-0.25in} \includegraphics[width=2.5in]{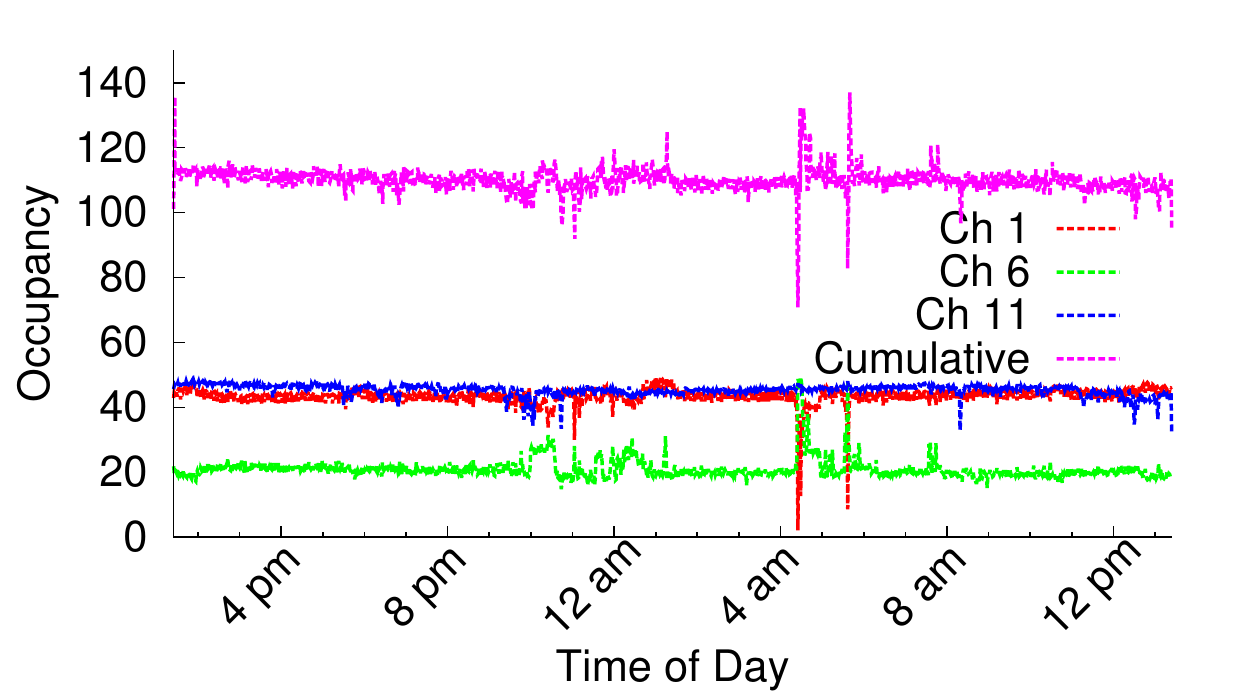}\\
\hspace{-0.25in} {(a) Home 1} & \hspace{-0.25in}{(b) Home 2} & \hspace{-0.25in}{(c) Home 3}
\end{tabular}
\begin{tabular}{ccc}
	\hspace{-0.25in} \includegraphics[width=2.5in]{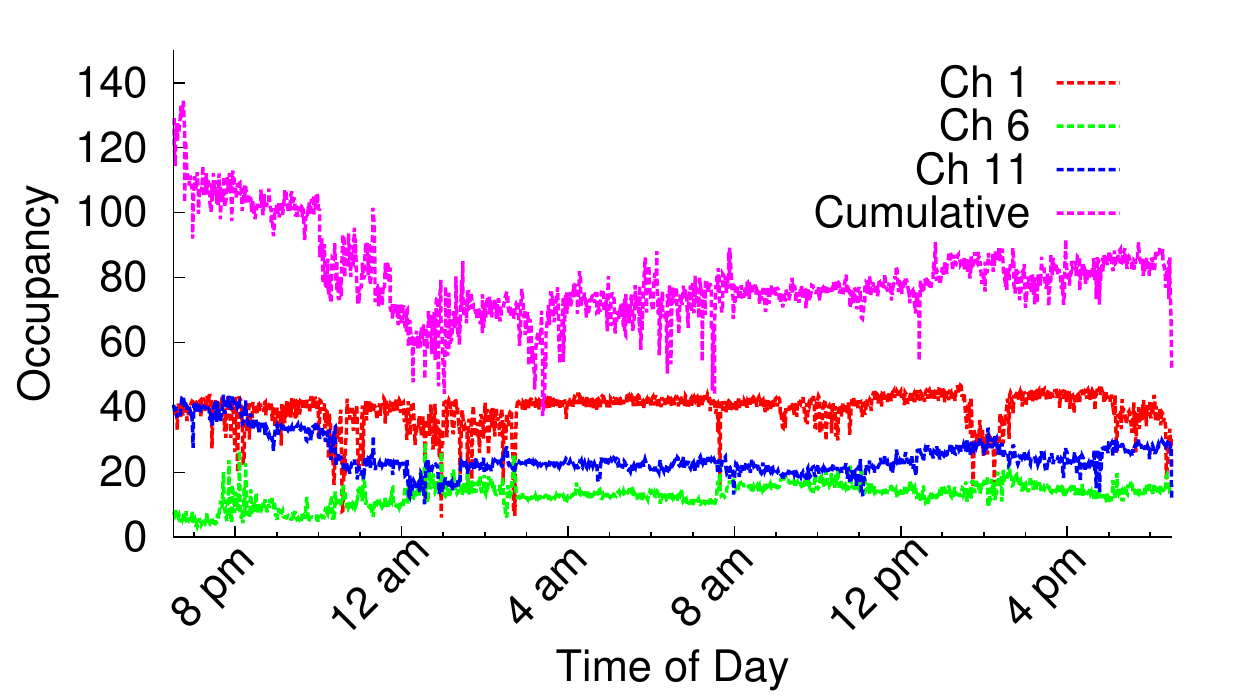} & \hspace{-0.25in} \includegraphics[width=2.5in]{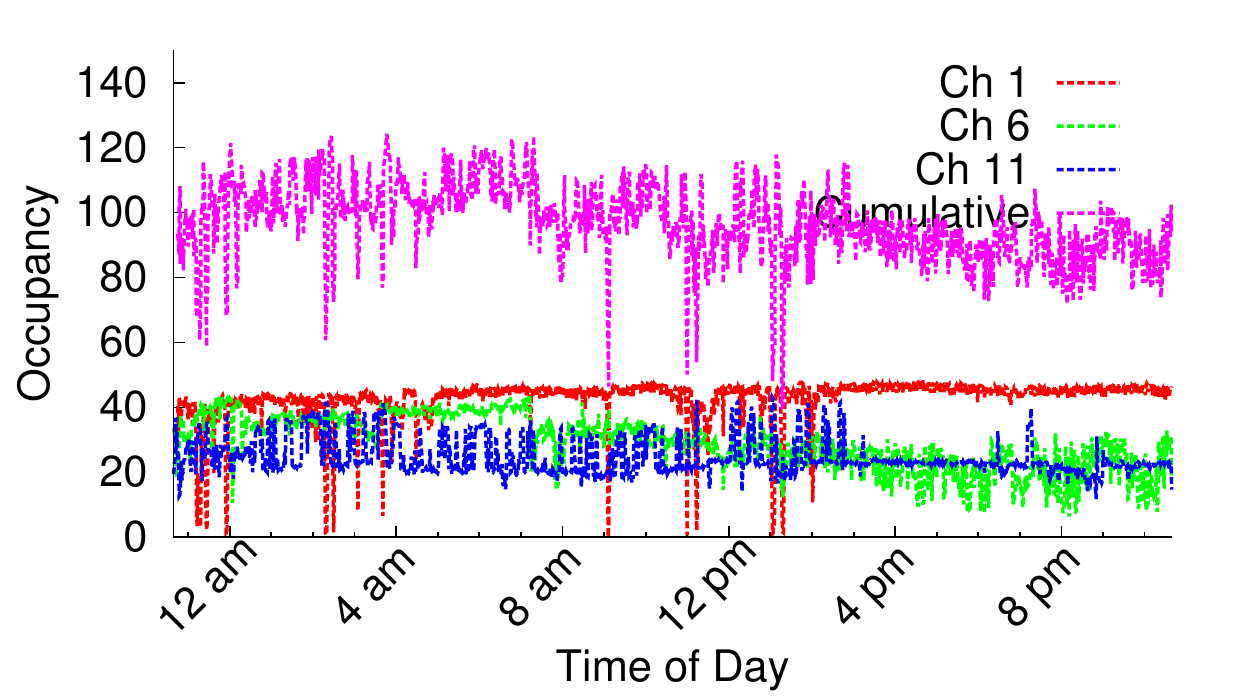} & \hspace{-0.25in} \includegraphics[width=2.5in]{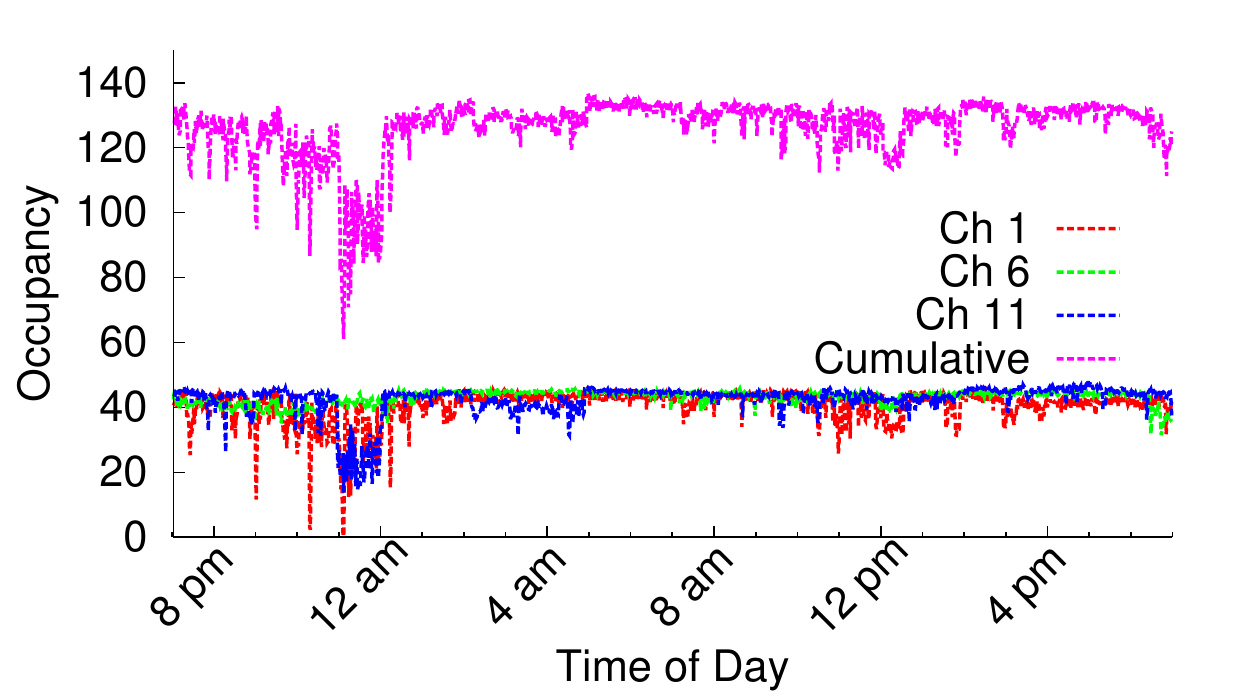}\\
\hspace{-0.25in} {(d) Home 4} & \hspace{-0.25in}{(e) Home 5} & \hspace{-0.25in}{(f) Home 6}
\end{tabular}
}
\vskip -0.15in
\caption{\footnotesize{\bf \name\ channel occupancies in our home deployments.} We see significant variation in the per-channel occupancy values across homes. This is because \name\ uses carrier sense that reduces its occupancy when the neighboring networks are loaded.  The cumulative occupancy, however, is high across time in all our home deployments. We note that, in principle, one can modify \name's algorithm to reduce the per-channel occupancy of the power traffic and keep the cumulative occupancy less than 100\%, which is sufficient for harvesting purposes.}
\label{fig:home}
\end{figure*}

\name's power-delivery efficiency depends on the traffic patterns of other Wi-Fi networks in the vicinity as well as the router's own client traffic, both of which can be unpredictable. To evaluate \name in practice, we deploy our system in six homes in a metropolitan area and measure its performance. Table~\ref{tab:home} summarizes the number of users, devices and other 2.4~GHz routers nearby in each of our deployments. We replace the router in each home with a \name router, and the occupants use it for normal Internet access for 24 hours. Our router uses the same SSID and authentication information as the original router, which we disconnect. We place our router within a few feet of the original router, with the exact location determined by user preferences. In all six deployments, we set our router to provide Internet connectivity on channel 1 and to transmit power packets on channels 1, 6, and 11 using the algorithm in \xref{sec:routerdesign}.  We stage our deployment over the period of a week---first two homes in Table~\ref{tab:home} over a weekend and the rest on weekdays.

We log the router's channel occupancy on each of the three Wi-Fi channels at a resolution of 60~s. \figref{fig:home} plots the occupancy values for each Wi-Fi channel over the 24-hour deployment duration. We also plot the cumulative occupancy across the channels. The figures show that:

\begin{Itemize}
\item
We see significant variation in  per-channel occupancy across homes. This is because when the load is high on  neighboring networks, our router scales back its transmissions on that channel and has lower channel occupancy. However, when the load on neighboring networks is low, the router occupies a larger fraction of the wireless channel. This is because \name\ uses carrier sense to enforce fairness with other Wi-Fi networks. 
\item
The cumulative occupancy is high over time in all our home deployments. Specifically, the mean cumulative occupancies for the six home deployments are in the 78-127\% range. We note that some of these occupancies are much greater than 100\%, which might not be necessary for power delivery. One can however reduce the per-channel rate of the power traffic based on the cumulative occupancy value to ensure that it is below 100\%. Our current system does not implement this feature.
\item
The users in homes 1--4 did not perceive any noticeable difference in their user experience. The user in home 5, however, noted a significant improvement in his page load times and better experience on streaming sites including Hulu, Amazon Prime and YouTube. This was primarily because home 5 originally was using a cheap low-grade router with worse specifications. A user in home 6 noted a slight deterioration in her YouTube viewing experience for a 30-minute duration. Our analysis showed that our router occupancy, including both client and power traffic, dipped during this duration. This points to external causes including interference from other devices in the environment.

\end{Itemize}

\begin{figure}[t!]
\vskip -0.1in
\centering
{\footnotesize
\begin{tabular}{c}
        \includegraphics[width=\columnwidth]{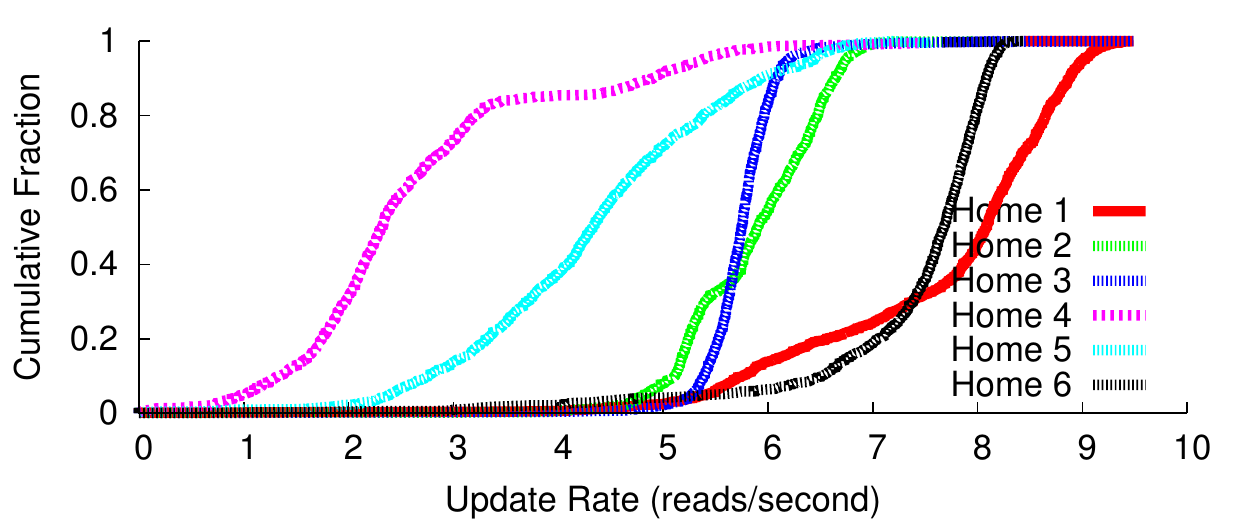}\\
\end{tabular}
}
\vskip -0.15in

\caption{{\bf Battery-free temperature sensor across homes.} The computed update rates ten feet away from our router, shows that we can deliver power via Wi-Fi with real-world network conditions.} 

\label{fig:home_dutycycle}
\vskip -0.09in
\end{figure}

Finally, \figref{fig:home_dutycycle} plots the CDFs of the computed update rates for our battery-free temperature sensor placed ten feet from the router in the homes. The plots show that we can successfully deliver power via Wi-Fi in real-world Wi-Fi network conditions.

\section{Related Work}
Wireless power delivery techniques can be primarily divided into two categories: near-field magnetic resonance/inductive coupling~\cite{sciencewitricity,chargecar,chargemat,magneticmimo, qispec} and RF power transmission systems. Of the two, RF power delivery is the truly long-range mechanism and hence we focus on the related work in the latter category.

Early RF power delivery systems were developed as part of RFID systems to harvest small amounts of power from a dedicated 900~MHz UHF RFID readers~\cite{wisp}. The power harvested from RFID signals has been used to operate microcontrollers~\cite{wisp}, LEDs~\cite{sample2012optical} and sensors such as accelerometers~\cite{wisp}, temperature sensors~\cite{wisp}, microphones~\cite{talla2013hybrid} and recently cameras~\cite{wisp_camera}. Our efforts on power deliver over Wi-Fi are complimentary to RFID systems. In principle, one can combine multiple ISM bands including 900~MHz, 2.4~GHz, and 5~GHz to design an optimal power delivery system. This paper takes a significant step towards this goal.

Recently, researchers have demonstrated the feasibility of harvesting small amounts of power from ambient TV~\cite{warp,tentzeris2013ambient,abc, multiband,mikeka2011dtv,kawahara2012sensprout} and cellular base station signals~\cite{visser2008ambient,parksGSMWARP} in the environment. While TV and cellular signals are stronger in outdoor environments, they are significantly attenuated indoors limiting the corresponding harvesting opportunities. The ability to power devices using Wi-Fi can augment the above capabilities and enable power harvesting indoors.

Researchers have also explored the feasibility of harvesting power in the 2.4~GHz ISM bands~\cite{valenta2014harvesting, andia2010design, volakis2012wifi,volakis2010planar, curty2005remotely, matt1, matt2, duke1, microwaveoven_harvesting, hagerty2004recycling}. These research efforts have demonstrated power harvesting from continuous wave transmissions\footnote{Continuous wave transmissions are special signals that have a constant amplitude and a single frequency tone.} and none have powered devices with existing Wi-Fi chipsets. Further,~\cite{duke1, microwaveoven_harvesting, hagerty2004recycling} harvest from incoming signals in excess of -5~dBm and can operate only in close proximity of the power source.~\cite{andia2010design, volakis2010planar} design a rectifier that outputs voltages around 100~mV for continuous wave transmissions at specific frequency tones. It is unclear how one may transform this into the 1.8--2.4~V required by microcontrollers, sensors and batteries.  \cite{curty2005remotely} discusses an IC implementation of a 2.45~GHz continuous-wave RFID tag. \cite{matt1} has recently analyzed the impact of the bursty nature of Wi-Fi traffic on the rectifier. It then optimizes the size of the rectifier's output capacitor based on Wi-Fi burstiness. However, similar to \cite{andia2010design, volakis2010planar}, this work is focused on rectifier design and does not power sensors and microcontrollers or recharge batteries. We also note that our work takes a different approach to the problem: We mask the burstiness in Wi-Fi traffic and instead create high cumulative channel occupancy at the router.~\cite{matt2} designs an efficient 2.4~GHz rectenna patch and battery charging solution but the rectenna is evaluated for continuous wave transmissions in an anechoic chamber and is not evaluated with Wi-Fi signals. In contrast, \name is the first power over Wi-Fi system that works with existing Wi-Fi chipsets and minimizes its impact on Wi-Fi performance.

Our work is also related to efforts from startups such as Ossia~\cite{ossia} and Wattup~\cite{wattup}. These efforts claim to deliver around 1 W of power at ranges of 15 feet and charge a mobile phone~\cite{wattup_article}. Back-of-the-envelope calculations however show that this requires continuous transmissions with an EIRP (equivalent isotropic radiated power) of 83.3~dBm (213~kW). This not only jams the Wi-Fi channel but also is 50,000 times higher power than that allowed by FCC regulations part 15 for point to multi-point links. In contrast, our system is designed to operate within the FCC limits and has minimal impact on Wi-Fi traffic. We note that in the event of an FCC exception to these startups, our multi-channel design can be used to deliver high power while having minimal effect on Wi-Fi performance.

Finally, recent work on Wi-Fi backscatter~\cite{wifibackscatter} enables low-power connectivity with existing Wi-Fi devices. Backscatter communication is order of magnitude more power-efficient than traditional radio communication and hence enables Wi-Fi connectivity without incurring Wi-Fi's power consumption. However, \cite{wifibackscatter} is focused on the communication mechanism and to the best of our knowledge, does not evaluate the feasibility of delivering power using Wi-Fi. Our work is complementary to~\cite{wifibackscatter} and can in principle be combined to achieve both power delivery and low-power connectivity using Wi-Fi devices.

\section{Discussions and Future Directions}

\begin{figure}[t!]
\centering
{\footnotesize
\begin{tabular}{c}
\includegraphics[width= 0.75\columnwidth]{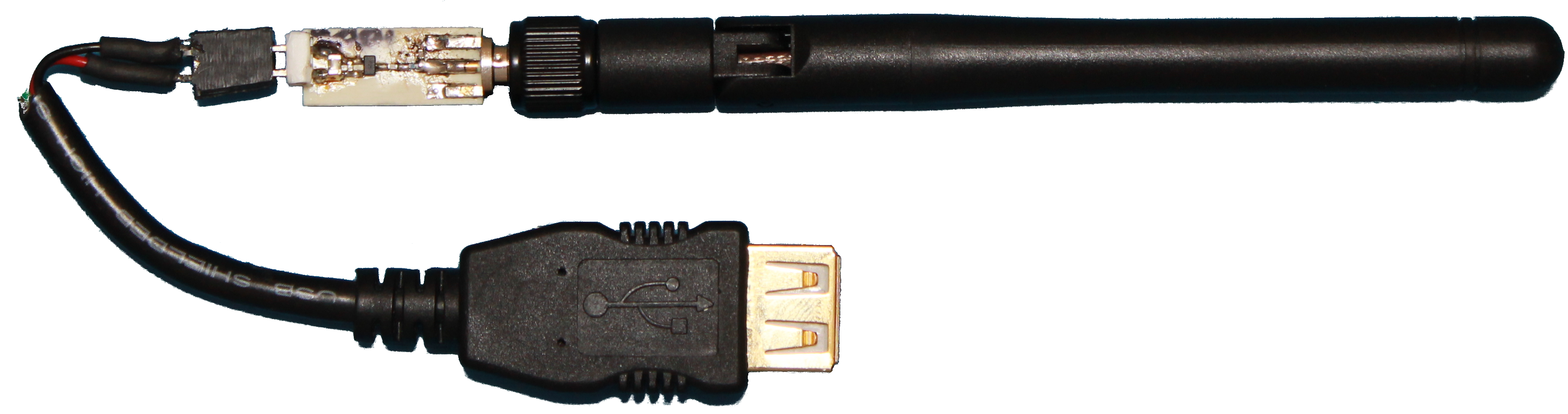}
\end{tabular}
}
\vskip -0.15in

\caption{{\bf Wi-Fi power via USB.} It consists of a 2~dBi Wi-Fi antenna attached to our harvester. Using this, we charge a Jawbone UP24 device in the vicinity of the \name\ router from a no-charge state to 41\% charged state in 2.5~hours.}

\label{fig:usb_charger}
\vskip -0.09in
\end{figure}

\vskip 0.05in\noindent{\it (a) Wi-Fi router as a charging hotspot.} In addition to powering custom temperature and camera sensors,  \name\ can transform the vicinity of a Wi-Fi router into a wireless charging hotspot for devices such as FitBit and Jawbone activity trackers. To demonstrate the feasibility of this, we design the general-purpose USB charger shown in \figref{fig:usb_charger}. It consists of a 2~dBi Wi-Fi antenna attached to a custom harvester that we optimize for higher input power values. We then connect our USB charger to a Jawbone UP24 device and place it 5-7~cm away from the \name\ router. We observe that the charger could supply an average current of 2.3~mA and charge the Jawbone UP24 battery from a no-charge state to 41\% charged-state in 2.5~hours. This demonstrates the potential of our approach. We are currently working on designs that would directly integrate our harvester with the antenna of the wearable device. Further, we are exploring the use of a custom battery charging solution, similar to those demonstrated in this paper, to achieve higher efficiencies and longer-distance wireless charging for these devices.

\vskip 0.05in\noindent{\it (b) \name with MIMO.} Our current implementation uses multiple antennas to transmit concurrently on different Wi-Fi channels. We could use MIMO techniques for transmitting to Wi-Fi clients on these antennas and use them for \name\ during the silent durations. We note that there is no fundamental tradeoff between MIMO and \name. Specifically, recent work~\cite{fullduplex-sigcomm12} has demonstrated that one can use a single antenna to transmit concurrently on adjacent Wi-Fi channels using circulators to cancel the signal leakages. Thus, we can, in principle, design Wi-Fi routers that concurrently transmit on adjacent Wi-Fi channels and also leverage MIMO~\cite{full-duplex-nsdi14}.

\vskip 0.05in\noindent{\it (c) Multiple \name\ routers.} In principle, multiple \name\ routers would have to time-multiplex their power traffic, thus reducing their cumulative channel occupancy and resulting in inefficient power delivery. Our solution is to allow \name\ routers to concurrently transmit their power packets. While this creates collisions between the power traffic, it is acceptable since our UDP broadcast packets do not need to be decoded by any specific client. As a result, the cumulative channel occupancy at each of the routers remains high. Implementing and evaluating this solution, however, is not in the scope of this paper.

\vskip 0.05in\noindent{\it (d) Security implications of \name.} As networks capable of delivering both power and data become prevalent, one can imagine a ``power denial-of-service'' (PDoS) attack in which a rogue device causes power starvation for other members of the network by generating signals designed to cause carrier sense events at the \name router.  This opens up interesting research opportunities for understanding the tradeoffs for security mechanisms that protect against such attacks in an efficient manner.

\vskip 0.05in\noindent{\it (e) Future clean-slate designs and \name.} We believe that our system is a general design for power delivery in the ISM bands. As Wi-Fi access and densities continue to grow in the ISM band, solutions that deteriorate Wi-Fi performance by jamming any specific frequency are not desirable. Our power delivery solution is integrated with the Wi-Fi protocol and hence can deliver power while having minimal impact on Wi-Fi traffic.  Future designs would generalize our multi-channel approach to operate across multiple ISM bands (e.g., 900~MHz, 2.4~GHz and 5~GHz). We believe that this paper takes a significant step towards that goal.

\section{Conclusion}
There is increasing interest in the Internet-of-Things where small computing sensors and mobile devices are embedded in everyday objects and environments. A key issue is how to power these devices as they become smaller and more numerous; plugging them in to provide power is inconvenient and is difficult at large scale.

We introduce a novel far-field power delivery system using existing Wi-Fi chipsets. We do so while minimizing the impact on  Wi-Fi network performance. We prototype the first battery-free temperature and camera sensors that are powered with Wi-Fi devices. We also demonstrate the feasibility of wirelessly recharging nickel--metal hydride and lithium-ion coin cell batteries. Finally, we deploy our system in multiple homes in a metropolitan area and demonstrate that \name\ can successfully deliver power via Wi-Fi with real-world Wi-Fi network conditions.

\balance


\let\oldthebibliography=\thebibliography
\let\endoldthebibliography=\endthebibliography
\renewenvironment{thebibliography}[1]{%
    \begin{oldthebibliography}{#1}%
      \setlength{\parskip}{0ex}%
      \setlength{\itemsep}{0ex}%
}%
{%
\end{oldthebibliography}%
}
{
\bibliographystyle{abbrv}
\bibliography{ourbib2}

}
\end{sloppypar}
\label{lastpage}
\end{document}